\documentclass[twocolumn,showpacs,preprintnumbers, superscriptaddress,eqsecnum,nofootinbib]{revtex4-1}
\usepackage{graphicx, fancybox}
\usepackage{amsmath,amssymb}
\usepackage[dvipdfm, colorlinks=true, pdfstartview=FitV, linkcolor=red, citecolor=blue, urlcolor=blue]{hyperref}
\usepackage{color}

\newcommand{\vect}[1]{\mathbf{#1}}

\newcommand{\im}{\mathrm{Im}}

\newcommand{\sgn}{\mathrm{sgn}}
\newcommand{\Slash}[1]{\ooalign{\hfil/\hfil\crcr$#1$}}

\newcommand{\vk}{\vect{k}}
\newcommand{\vx}{\vect{x}}
\newcommand{\vy}{\vect{y}}

\newcommand{\mb}{m_b}
\newcommand{\mf}{m_f}
\newcommand{\me}{m_e}
\newcommand{\mg}{m_\gamma}
\newcommand{\zetae}{\zeta_e}
\newcommand{\comment}[1]{}

\newcommand{\jind}{j_{\mathrm {ind}}}
\newcommand{\etaind}{\eta_{\mathrm {ind}}}

\newcommand{\Tc}{{\mathrm T}_C}
\newcommand{\R}{R}
\newcommand{\A}{A}
\newcommand{\cp}{g}
\newcommand{\nf}{n_F}
\newcommand{\nb}{n_B}
\newcommand{\VC}{ \varXi }
\newcommand{\V}{\tilde{\varGamma}}
\newcommand{\GF}{\tilde{G}}
\newcommand{\thetac}{\theta_C}

\begin{document}

\preprint{RIKEN-QHP-23, RIKEN-MP-44}

\title{Off-diagonal kinetic theory in ultrasoft momentum region at high temperature}

\author{Daisuke Satow}
\email{d-sato@ruby.scphys.kyoto-u.ac.jp}
\affiliation{Department of Physics, Faculty of Science, Kyoto University, Kitashirakawa Oiwakecho, Sakyo-ku, Kyoto 606-8502, Japan}
\affiliation{Institut de Physique Th\'eorique, Orme des Merisiers batiment 774 Point courrier 136 CEA/DSM/IPhT, CEA/Saclay, F-91191 Gif-sur-Yvette Cedex, France}

\author{Yoshimasa Hidaka}
\email{hidaka@riken.jp}
\affiliation{Quantum Hadron Physics Laboratory, RIKEN Nishina Center, Saitama 351-0198, Japan}

\begin{abstract}
We derive a linearized kinetic equation for fermionic excitations with an ultrasoft momentum, $\cp^2T$, from the Kadanoff-Baym equation
in a Yukawa model and quantum electrodynamics (QED) at extremely high $T$, where $g$ is the coupling constant.
We show that this equation is equivalent to the self-consistent equation in the
resummed perturbation theory used in the analysis of the fermion spectrum with the ultrasoft momentum. 
Furthermore, we derive the equation that determines the $n$-point function 
with external lines for a pair of fermions and $(n-2)$ bosons with ultrasoft momenta in QED.
\end{abstract}

\date{\today}

\maketitle

\section{Introduction}
\label{sec:intro}
Relativistic plasmas at extremely high temperature $T$ such as electron and/or quark-gluon plasma are expected to be realized in the early universe and the experiments of heavy ion collisions at Relativistic Heavy Ion Collider and Large Hadron Collider.
These systems have multi-energy scale at weak coupling ($\cp\ll 1$)
even at so high $T$ that all mass scale in the Lagrangian is negligible;
e.g, typical energy of particles is of order $T$,
while the energy of the bosonic and fermionic collective excitations called plasmon~\cite{plasmon} and plasmino~\cite{plasmino} is of order $\cp T$, and
the fermion damping rate in gauge theories is of order $\cp^2T$~\cite{electron-damping}. 
These energy scales, $T$, $\cp T$, and $\cp^2T$, are called hard, soft, and ultrasoft scales, respectively.

\begin{figure*}[t]
\begin{center}
\includegraphics[width=0.70\textwidth]{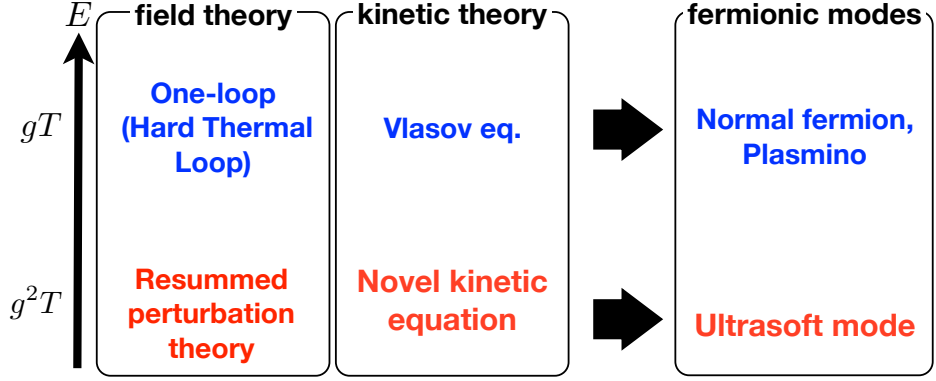}
\caption{The correspondence between the resumed perturbation theory and the kinetic theory, and the fermionic modes obtained from these theories.
The vertical axis denote the energy scale.}
\label{fig:intro}
\end{center}
\end{figure*}

Unlike perturbation theory in the vacuum, the loop expansion at finite temperature
does not correspond to the coupling nor $\hbar$ expansions, and
the scale dependent expansion is necessary (Fig.~\ref{fig:intro}).
When the energy scale is $\cp T$, the one-loop approximation obtained by integrating out hard internal momenta,
called hard thermal loop (HTL) approximation~\cite{HTL}, is reliable, and the approximation establishes the HTL effective theory~\cite{HTL-resum}.
If the energy is of order or much less than $\cp^2T$, some resummation\footnote{This resummation~\cite{susy-kinetic,transport-perturbation,transport-kinetic,transport-2PI,blaizot-ultrasoft,lebedev,QED} is different from the well-known HTL resummation,
which is the resummation for the contribution from soft-internal momenta~\cite{HTL-resum}. } is necessary~\cite{susy-kinetic,transport-perturbation,transport-kinetic,transport-2PI,blaizot-ultrasoft,lebedev,QED}.
Here let us focus on the analysis for the fermion propagator with an ultrasoft momentum. The resummation scheme used in the analysis of the fermion propagator consist of the following procedures~\cite{lebedev,QED}:
\begin{enumerate}
\item Resumming the thermal mass and the decay width of the hard particles.
\item Summing all the ladder diagrams (in gauge theory).
\end{enumerate} 
By analyzing the fermion propagator using the resummed perturbation theory, a novel fermionic excitation was found~\cite{lebedev,QED}.

Each perturbation scheme with soft and ultrasoft momenta can be interpreted as different kinetic equations.
When the energy is soft, the effect of collisions  is negligible and hence the collisionless  kinetic equation  called Vlasov equation is valid~\cite{blaizot-HTL}.
This equation corresponds to the HTL approximation~\cite{HTL} introduced above.
When the energy is of order or less than $\cp^2T$, the effects of the interaction among the hard particles are no longer negligible in general, so it is necessary to take into account their effects.
In fact, the analysis on the gluon self-energy with an ultrasoft momentum should be performed with the resummed perturbation theory including the effect of collisions, and it was done by constructing the linearized Boltzmann equation~\cite{blaizot-ultrasoft}.
The computation of transport coefficients, whose energy scale is much smaller than $\cp^2T$, is also performed with the resummed perturbation theory~\cite{transport-perturbation}, the linearized Boltzmann equation~\cite{transport-kinetic}, and the two-particle irreducible formalism~\cite{transport-2PI}.
These methods produce the same result in the leading order of the coupling constant.
Using the correspondence between the perturbation theory and the linearized kinetic theory, the self-consistent equation in the resummed perturbation theory is interpreted with the language of the kinetic theory.

In this paper, we derive a generalized and linearized kinetic equation for fermionic excitations with an ultrasoft momentum, which we will call ``off-diagonal'' kinetic equation later,  in the Yukawa model and QED, while the Boltzmann equation discussed above is employed for bosonic excitations.
Our equation is systematically derived from the Kadanoff-Baym equation, 
and is equivalent to the self-consistent equation in the
resummed perturbation theory~\cite{QED, lebedev} used in the analysis of the fermion propagator.
The derivation helps us to establish the foundation of the resummed perturbation scheme.
The kinetic equation will also give us the kinetic interpretation of the resummation scheme.
Furthermore, we also discuss the procedure of analyzing the higher point functions not only two-point function of the fermion in QED.

This paper is organized as follows:
Section~\ref{sec:yukawa} is devoted to the derivation of the generalized and linearized kinetic equation and the discussion on the kinetic interpretation of the self-consistent equation in the resummed perturbation theory in the Yukawa model, which is the simplest fermion-boson system.
In Sec.~\ref{sec:QED}, a similar analyses in QED is done in the Coulomb gauge.
We make concluding remarks and summarize our results in Sec.~\ref{sec:summary}.
In Appendix~\ref{app:analyticity}, we evaluate the structure of the induced source terms.
We analyze the Kadanoff-Baym equation that would yield the linearized kinetic equation in the diagonal case in Appendix~\ref{app:diagonal}.
Appendix~\ref{app:K0} is devoted to showing that the temporal component of $K^\mu(x,y)$ introduced in Sec.~\ref{sec:QED} can be neglected.
We show that the result obtained in Sec.~\ref{sec:QED} is valid also in the temporal gauge, which is the gauge-fixing of the original paper~\cite{lebedev} of the resummed perturbation theory, in Appendix~\ref{app:temporal}.

\section{Kinetic equation in Yukawa model}
\label{sec:yukawa}

\begin{figure}[t]
\begin{center}
\includegraphics[width=0.45\textwidth]{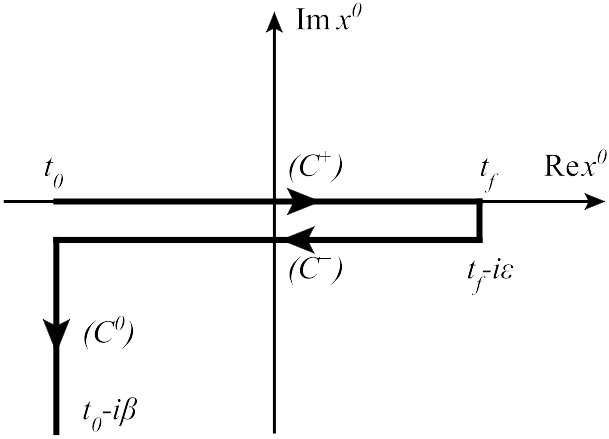}
\caption{The contour path $C$ in the complex $x^0$ plane.}
\label{fig:contour}
\end{center}
\end{figure}

In this section,  we derive a novel linearized kinetic equation from the Kadanoff-Baym equation in the Yukawa model, which is the simplest model for  fermion-boson systems.
We will find the vertex correction is negligible, which makes the analysis simpler than that in gauge theories.
Next, we show that the kinetic equation is equivalent to the self-consistent equation in the resummed perturbation theory~\cite{susy-kinetic,QED}, and discuss the interpretation of the self-consistent equation using the correspondence to the kinetic theory.

\subsection{Derivation of the kinetic equation}
\label{ssc:yukawa-derivation}
Throughout this paper, we work in the closed-time-path formalism~\cite{lebellac,blaizot-review}.
We perform the derivation of the kinetic equation in a similar way used in~\cite{blaizot-HTL,blaizot-review}
by applying the gradient expansion to the Kadanoff-Baym equation~\cite{kadanoff-baym} and taking into account the interaction effect among the hard particles in the leading order.

Let us consider the following situation to analyze the fermionic ultrasoft excitation:
Before the initial time $t_0$, the system is at equilibrium with a temperature $T$.
Then, a (anti-) fermionic external source $\eta(x)$ ($\overline{\eta}(x)$)
and a scalar external source $j(x)$ are switched on. 
As a result, the system becomes nonequilibrium. 
We will consider the case that $j(x)$ and $\overline{\eta}(x)$ vanish and $\eta(x)$ is so weak that the system is very close to the equilibrium, i.e.,  the linear response regime. 
Concretely, we will retain only the terms in the linear order of the fermionic average field $\varPsi$ in the fermionic induced source, which will be introduced later.

Let us consider the generating functional in the closed time formalism~\cite{blaizot-review},
\begin{equation} 
\label{eq:yukawa-Z} 
Z[j,\eta,\overline{\eta}]=\int {\cal D}{\phi}\,{\cal D}\overline{\psi}\,{\cal D}\psi \,e^{iS},
\end{equation}
with
\begin{equation}
S=\int_C d^4x
\bigl[  {\cal L}[\phi, \psi, \overline{\psi}]-(j\phi+\overline{\psi}\eta+\overline{\eta}\psi)\bigr],
\end{equation}
where $\phi$ and $\psi$ are the scalar and the fermion fields.
The space-time integral is defined as $\int_C d^4x\equiv\int_C dx^0 \int d^3 \vx$, where $\int_C dx^0$ is the complex-time integral along the contour $C=C^+\cup C^-\cup C^0$ in Fig.~\ref{fig:contour}.
 We will take $t_0\to -\infty$ and $t_f\to \infty$ to factorize out the contribution from the path $C^0$.
The Lagrangian in the Yukawa model with the massless fermion and boson has the form, 
\begin{align}
{\cal L}[\phi, \psi, \overline{\psi}]= \frac{1}{2}(\partial^\mu\phi)^2+\overline{\psi}i(\Slash{\partial}+i\cp\phi)\psi,
\end{align}
where we did not take into account the self-interaction of $\phi$ such as $\lambda\phi^4$ for simplicity.
By performing an infinitesimal variation with respect to $\phi$ or $\psi$ in Eq.~(\ref{eq:yukawa-Z}), we obtain the following equations of motion:
\begin{align}
\label{eq:yukawa-eom-meanfield-fermion}
i\Slash{D}_x[\varPhi]\varPsi(x)&= \eta(x)+\etaind(x), \\ 
\label{eq:yukawa-eom-meanfield-boson}
-\partial^2\varPhi(x)-\cp{\overline{\varPsi}} \varPsi (x)&=j(x)+\jind(x), 
\end{align}
where $\varPhi\equiv \langle\phi\rangle$ ($\varPsi\equiv\langle \psi\rangle$) is the expectation value of the scalar (fermion) field, and
$\Slash{D}_x[\varPhi]\equiv\Slash{\partial}_x+i\cp\varPhi(x)$.
Here the expectation value for an operator $\mathcal{O}$ is defined as
\begin{equation}
\begin{split}
\langle\mathcal{O}\rangle \equiv \frac{1}{Z}\int {\cal D}{\phi}\,{\cal D}\overline{\psi}\,{\cal D}\psi\, e^{iS}  \mathcal{O}.
\end{split}
\end{equation}
$\etaind(x)\equiv  \cp\langle\phi(x) \psi(x)\rangle_c$ ($\jind(x)\equiv \cp\langle\overline{\psi}(x)\psi(x)\rangle_c$) is the fermionic (scalar) induced source, 
and the subscript $c$ denotes  ``connected,'' i.e., $\langle\phi(x) \psi(x)\rangle_c\equiv \langle\phi(x) \psi(x)\rangle-\varPhi(x)\varPsi(x)$.

By differentiating Eq.~(\ref{eq:yukawa-eom-meanfield-fermion}) with respect to $j(y)$ and Eq.~(\ref{eq:yukawa-eom-meanfield-boson}) with respect to $\overline{\eta}(y)$, we obtain
\begin{align}
\label{eq:yukawa-fluctuation-fermion}
 &i\Slash{D}_x[\varPhi] K(x,y)-gD(x,y)\varPsi(x)=i\frac{\delta \etaind(x)}{\delta j(y)},\\
 \notag
&\partial^2_xK(y,x)-\cp(\overline{\varPsi}(x)\langle\psi(y)\psi(x)\rangle_c+S(y,x)\varPsi(x))\\
\label{eq:yukawa-fluctuation-boson}
&\quad=i\frac{\delta \jind(x)}{\delta \overline{\eta}(y)}.
\end{align}
Here we have introduced the following propagators:
\begin{align}
D(x,y)&\equiv  \langle\Tc\phi(x)\phi(y)\rangle_c = i\frac{\delta\varPhi(x)}{\delta j(y)},\\
S(x,y)&\equiv \langle\Tc\psi(x)\overline{\psi}(y)\rangle_c=i\frac{\delta \overline{\varPsi}(y)}{\delta\overline{\eta}(x)},\\
K(x,y) &\equiv \langle\Tc\psi(x)\phi(y)\rangle_c = i\frac{\delta \varPsi(x)}{\delta j(y)}=i\frac{\delta \varPhi(y)}{\delta\overline{\eta}(x)},
\end{align}
where $\Tc$ means the path ordering  on the complex-time path $C$;
explicitly, 
\begin{align}
\begin{split}
D(x,y)&= \thetac(x^0, y^0)D^>(x,y)+\thetac(y^0, x^0)D^<(x,y),\\
S(x,y)&=  \thetac(x^0, y^0)S^>(x,y)-\thetac(y^0, x^0)S^<(x,y),\\
K(x,y)&= \thetac(x^0, y^0)K^>(x,y)+\thetac(y^0, x^0)K^<(x,y),
\end{split}
\end{align}
with 
\begin{align}
D^>(x,y)&\equiv \langle\phi(x) \phi(y)\rangle_c,\\
D^<(x,y)&\equiv \langle\phi(y)\phi(x)\rangle_c,\\
S^>(x,y)&\equiv \langle\psi(x) \overline{\psi}(y)\rangle_c,\\
S^<(x,y)&\equiv \langle\overline{\psi}(y)\psi(x)\rangle_c,\\
K^>(x,y)&\equiv  \langle\psi(x)\phi(y)\rangle_c,\\
K^<(x,y)&\equiv \langle\phi(y)\psi(x)\rangle_c,
\end{align}
and $\thetac(x,y)$ being the step-function along the path $C$.
In the approximations introduced later, we can see that $K^>(x,y)$ and $K^<(x,y)$ coincide,  which can be checked by $K^{\R}(x,y)\equiv i\theta(x^0, y^0)[K^>(x,y)-K^< (x,y)] \simeq0$. 
For this reason, we simply write these two functions as $K(x,y)$ from now on.
We call $K(x,y)$ ``{\it{off-diagonal propagator}},'' which mixes the fermion and boson, 
while we call $D(x,y)$ and $S(x,y)$ ``{\it{diagonal propagators}}.''
As will be seen in Sec.~\ref{ssc:yukawa-correspond},
in the calculation of the ultrasoft fermion self-energy,
the off-diagonal propagator is more relevant than other diagonal ones.

By setting $x^0\in C^+$ and $y^0\in C^-$ in  Eqs.~(\ref{eq:yukawa-fluctuation-fermion}) and (\ref{eq:yukawa-fluctuation-boson}), we obtain 
\begin{align}
\label{eq:yukawa-fermion1}
 &i\Slash{D}_x[\varPhi] K(x,y)-\cp D^<(x,y)\varPsi(x)
 =i\frac{\delta \etaind(x)}{\delta j(y)},\\
 \notag
&-\partial^2_yK(x,y)+\cp(\overline{\varPsi}(y)\langle \psi(y)\psi(x)\rangle_c+S^<(x,y)\varPsi(y))\\
\label{eq:yukawa-boson1}
&\quad=i\frac{\delta \jind(y)}{\delta \overline{\eta}(x)}.
\end{align}
Here we have interchanged $x$ and $y$ in the second equation.
Let us evaluate the right-hand side of Eqs.~(\ref{eq:yukawa-fermion1}) and (\ref{eq:yukawa-boson1}) using the chain rule:
\begin{align}
&\frac{\delta \etaind(x)}{\delta j(y)} \notag\\
&\quad=\int_C d^4 z \left(\frac{\delta \etaind(x)}{\delta \varPsi(z)}\frac{\delta \varPsi(z)}{\delta j(y)}
+\frac{\delta \etaind(x)}{\delta \varPhi(z)}\frac{\delta \varPhi(z)}{\delta j(y)}\right) \notag\\
\label{eq:yukawa-fermion2}
&\quad=\int_C d^4z (\varSigma(x,z) K(z,y)+ \VC(x,z)D(z,y)),\\
\notag
&\frac{\partial \jind(y)}{\partial \overline{\eta}(x)} \\
\notag
&\quad=\int_C d^4 z \left(\frac{\delta \varPhi(z)}{\delta \overline{\eta}(x)} \frac{\delta \jind(y)}{\delta \varPhi(z)}
+\frac{\delta \overline{\varPsi} (z)}{\delta \overline{\eta}(x)}\frac{\delta \jind(y)}{\delta \overline{\varPsi} (z)}\right)\\
\label{eq:yukawa-boson2}
&\quad=\int_C d^4z (\varPi(y,z) K(x,z)+ S(x,z)\VC(z,y)).
\end{align} 
Here we have dropped $(\delta\etaind /\delta \overline{\varPsi} )(\delta \overline{\varPsi} /\delta j)$  and $(\delta \jind/\delta \varPsi) (\delta \varPsi /\delta \overline{\eta})$ since they contain more than one fermionic average field.
We have also used $\varSigma(x,y)\equiv-i\delta \etaind(x)/\delta\varPsi(y)$ and $\varPi(x,y)\equiv -i \delta \jind(x)/\delta \varPhi(y)=\varPi(y,x)$, where $\varSigma$ ($\varPi$) is the fermion (scalar) self-energy~\cite{blaizot-review,blaizot-ultrasoft}.
We also introduced the {\it off-diagonal} self-energy, $\VC(x,y)\equiv-i\delta \etaind(x)/\delta \varPhi(y)=-i\delta \jind(y)/\delta \overline{\varPsi} (x)$.

The self-energies are decomposed for arbitrary $x^0$ and $y^0$ on the time path $C$: 
\begin{align}
\varPi(x,y)&=\thetac(x^0, y^0)\varPi^>(x,y)+\thetac(y^0, x^0)\varPi^<(x,y),\\
\varSigma(x,y)&=\thetac(x^0, y^0)\varSigma^>(x,y)-\thetac(y^0, x^0)\varSigma^<(x,y),\\
\VC(x,y)&=\thetac(x^0, y^0)\VC^{>}(x,y) +\thetac(y^0, x^0)\VC^{<}(x,y).
\end{align}
We have not taken into account contact terms, which is negligible in the leading order as will be seen later.  
From Eqs.~(\ref{eq:app-ana-1}), (\ref{eq:app-ana-2}), (\ref{eq:app-ana-4}), and (\ref{eq:app-ana-5}) in the Appendix, 
we can rewrite Eqs.~(\ref{eq:yukawa-fermion1}) and (\ref{eq:yukawa-boson1}) in terms of real time integral instead of that on the complex-time-path:
\begin{align}
\notag
&i\Slash{D}_x[\varPhi] K(x,y)-\cp D^<(x,y)\varPsi(x)\\
\label{eq:yukawa-fermion3}
&\quad=\int ^\infty_{-\infty}d^4z (\varSigma^\R(x,z) K(z,y)+ \VC^\R(x,z)D^<(z,y)),\\
\notag
&-\partial^2_yK(x,y)+\cp S^<(x,y)\varPsi(y)\\
\label{eq:yukawa-boson3}
&\quad=\int ^\infty_{-\infty} d^4z (\varPi^\A(z,y) K(x,z)- S^<(x,z)\VC^\R(z,y)),
\end{align}
 where the retarded fermion (advanced scalar) self-energy $\varSigma^\R(x,y)\equiv i \theta(x^0, y^0)[\varSigma^>(x,y)+\varSigma^<(x,y)]$ ($\varPi^A(x,y)\equiv -i\theta(y^0, x^0)[\varPi^>(x,y)-\varPi^<(x,y)] $) and the retarded off-diagonal self-energy $\VC^{R}(x,z)\equiv i\theta(x^0, z^0)[\VC^>(x,z)-\VC^< (x,z)] $ have been introduced, and the limits $t_0\rightarrow -\infty$ and $t_f\rightarrow\infty$ have been taken.
Here we  dropped $\overline{\varPsi}(y)\langle\psi(y)\psi(x) \rangle_c$ because $\langle\psi(y)\psi(x) \rangle_c$ contains more than one $\varPsi$.
Equations~(\ref{eq:yukawa-fermion3}) and (\ref{eq:yukawa-boson3}) are the Kadanoff-Baym equations from which the kinetic equation is derived.

Let us introduce the Wigner transformation to derive the kinetic equation,
which is defined as 
\begin{align}
f(k,X)\equiv\int d^4s e^{ik\cdot s}f\left(X+\frac{s}{2}, X-\frac{s}{2}\right),
\end{align}
where $s\equiv x-y$, $X\equiv(x+y)/2$, and $f(x,y)$ is an arbitrary function.
After performing the Wigner transformation, Eqs.~(\ref{eq:yukawa-fermion3}) and (\ref{eq:yukawa-boson3}) become
\begin{align}
\notag
&\left(-i\Slash{k}+\frac{\Slash{\partial}_X}{2}+i\cp\varPhi(X)\right) K(k,X)+i\cp D^<(k,X)\varPsi(X)\\
\label{eq:yukawa-fermion4}
&\quad=i (-\varSigma^\R(k,X) K(k,X)-\VC^\R(k,X)D^<(k,X)),\\
\notag
&(k^2-ik\cdot\partial_X)K(k,X)+\cp S^<(k,X)\varPsi(X)\\
\label{eq:yukawa-boson4}
&\quad=\varPi^\A(k,X) K(k,X)- S^<(k,X)\VC^\R(k,X).
\end{align}
Here we have used the following transformation law under the Wigner transformation,
\begin{align}
\nonumber
f(x)g(x,y)&\rightarrow f(X)g(k,X)\\
&-\frac{i}{2}(\partial_X f)\cdot (\partial_k g)+..., \\
\nonumber
f(y)g(x,y)&\rightarrow f(X)g(k,X)\\
&+\frac{i}{2}(\partial_X f)\cdot (\partial_k g)+..., \\
\nonumber
\int^\infty_{-\infty} d^4z g(x,z)h(z,y)&\rightarrow g(k,X)h(k,X)\\
&+\frac{i}{2}\{g ,h \}_{\text {P. B.}}+... ,
\end{align}
where $\{g, h\}_{\text {P. B.}}\equiv\partial_k g\cdot \partial_X h-\partial_X g\cdot \partial_k h$ is the Poisson bracket, and neglected higher-order terms that contain $\partial_X$ since we focus on the case that the inhomogeneity of the average field is $\partial_X\sim \cp^2 T$, while a typical magnitude of $k$ is of order $T$.
This expansion is called gradient expansion~\cite{blaizot-HTL, blaizot-ultrasoft,blaizot-review}.
We retained the second terms in the left-hand sides of Eqs.~(\ref{eq:yukawa-fermion4}) and (\ref{eq:yukawa-boson4}) because the first terms, which seem to be the leading terms in the gradient expansion, will cancel out in the next manipulation.

By multiplying Eq.~(\ref{eq:yukawa-fermion4}) by $(-i\Slash{k}+\Slash{\partial}_X/2+i\cp\varPhi(X)+i\varSigma^\R(k,X) )$, adding Eq.~(\ref{eq:yukawa-boson4}), and setting $\varPhi=0$, we get
\begin{align}
\label{eq:yukawa-A}
\begin{split}
&(2ik\cdot\partial_X -\{\Slash{k}, \varSigma^\R(k,X)\}+\varPi^\A(k,X)) K(k,X)\\
&= \cp(\Slash{k}D^<(k,X)+S^<(k,X))\V(k,X).
\end{split}
\end{align}
Here we have introduced $\cp\V(k,X)\equiv \cp\varPsi(X)+\VC^\R(k,X)$.
We have neglected higher order terms of order $\cp^4T^2K$ and $\cp^3 T^{-1} \V$.
In the leading order, the coupling dependence in $D^{<}(k)$ and $S^{<}(k)$ is negligible, so that $D^{<}(k)$ and $S^{<}(k)$ are replaced by the propagators at equilibrium and free limit ($\cp=0$):
\begin{align}
D^{0<}(k)&=\rho^0(k)\nb(k^0), \\
S^{0<}(k)&= \Slash{k}\rho^0(k)\nf(k^0),
\end{align}
where $\rho^0(k)$ is the free spectral function given by
\begin{align}
\begin{split}
\rho^0(k)&\equiv 2\pi\sgn(k^0)\delta(k^2)\\
&=\frac{2\pi}{2|\vk|}(\delta(k^0-|\vk|)-\delta(k^0+|\vk|)), \label{eq:spectralFunction}
\end{split}
\end{align}
and $\nb(k^0)\equiv (e^{k^0/T}-1)^{-1}$ ($\nf(k^0)\equiv(e^{k^0/T}+1)^{-1}$) is the boson (fermion) distribution function at equilibrium.
We note that though the massless condition $k^2=0$ appears in Eq.~(\ref{eq:spectralFunction}) in the present approximation, $k^2$ is expected to be of order $\cp^2T^2$ if one takes into account the interaction at equilibrium.
For this reason, we will use the order estimate $k^2\sim \cp^2T^2$.
We also note that $K$ can not be replaced by that at equilibrium since $K$ vanishes at equilibrium.

\begin{figure}[t]
\begin{center}
\includegraphics[width=0.25\textwidth]{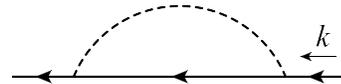}
\caption{The fermion retarded self-energy $\varSigma^{R {\mathrm {(eq)}}}(k)$ in the leading order.
The solid line is the fermion propagator and the dashed line is the scalar (the case of the Yukawa model) or photon propagator (the case of QED).
We note that the photon propagator in the case of QED should be replaced with the HTL-resummed one~\cite{HTL-resum} in this figure.}
\label{fig:fermion-selfenergy}
\end{center}
\end{figure}

\begin{figure}[t]
\begin{center}
\includegraphics[width=0.3\textwidth]{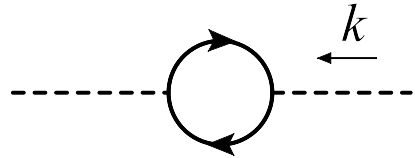}
\caption{The boson advanced self-energy $\varPi^{\A {\mathrm {(eq)}}}(k)$ in the leading order.
The notations are the same as Fig.~\ref{fig:fermion-selfenergy}.}
\label{fig:boson-selfenergy}
\end{center}
\end{figure}

We see that $k^2$ terms in the left-hand side of Eq.~(\ref{eq:yukawa-A}) were canceled out and $k\cdot\partial_X\sim \cp^2T^2$ term remains.
Thus, we can neglect the terms much smaller than $\cp^2T^2 K$ in the calculation of the leading order.
Following this line, the diagonal self-energies are replaced by those at equilibrium in the leading order, whose diagrams are shown in Figs.~\ref{fig:fermion-selfenergy} and \ref{fig:boson-selfenergy}:
\begin{align}
\{\Slash{k}, \varSigma^{\R {\mathrm {(eq)}}}(k)\}&= \mf^2,\\ 
\varPi^{\A {\mathrm {(eq)}}}(k)&= \mb^2, 
\end{align}
where $\mf\equiv \cp T/(2\sqrt{2})$ and $\mb\equiv \cp T/\sqrt{6}$ are  asymptotic thermal masses for
the fermion and the scalar boson in the leading order~\cite{scalar}.
Note that the imaginary parts of the self-energies $\sim g^4T^2 \ln( 1/g)$ and momentum dependence are negligible due to higher order of the coupling constant.
We have used the on-shell condition, $k^2\simeq 0$, which will be verified later.

\begin{figure}[t]
\begin{center}
\includegraphics[width=0.27\textwidth]{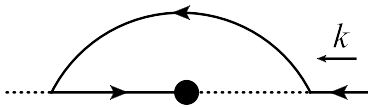} 
\caption{The off-diagonal self-energy $\VC(k, X)$ in the leading order.
The propagator that is composed of the solid line and the dashed line with the black blob is the off-diagonal propagator. 
The other notations are the same as Fig.~\ref{fig:fermion-selfenergy}.}
\label{fig:vertex-correction}
\end{center}
\end{figure}

The same line as the diagonal self-energies case justifies substituting the off-diagonal self-energy in the leading order, shown in Fig.~\ref{fig:vertex-correction} diagrammatically.
The expression is given by
\begin{align}
\label{eq:yukawa-vertex-correction}
\VC^\R(k,X)=& -\cp^2\int\frac{d^4k'}{(2\pi)^4}S^{0\R}(k+k')K(k',X).
\end{align}
Here $S^{0\R}(k)\equiv-\Slash{k}/((k_0+i\epsilon)^2-\vk^2)$ is the free fermion retarded propagator at equilibrium.
We note that the self-energies can not be neglected unlike $\partial_X\sim\cp T$ case\footnote{This is because $k\cdot\partial_X K\sim \cp T^2K\gg \{\Slash{k}, \varSigma\}K$, $\varPi K$, $ T^{-1}\VC\sim \cp^2T^2K$.}~\cite{blaizot-HTL}, because $\{\Slash{k}, \varSigma\} K$, $\varPi K$, $ T^{-1}\VC$ is the same order as $ \cp^2 T^2K$.

\begin{figure}[t]
\begin{center}
\includegraphics[width=0.45\textwidth]{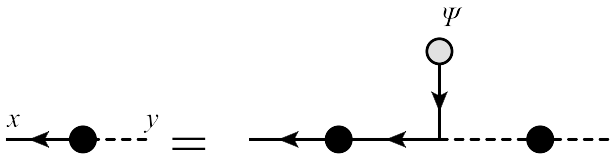}
\caption{The schematic figure of the off-diagonal propagator $K(x,y)$ in the Yukawa model in the leading order in the linear response regime.
The solid (dashed) line with black blob is the resummed fermion (boson) propagator that contains the information on the fermion (boson) self-energy $\varSigma$ ($\varPi$).
The gray blob represents the fermionic average field $\varPsi$.
}
\label{fig:yukawa-K}
\end{center}
\end{figure}

Using these expressions, Eq.~(\ref{eq:yukawa-A}) becomes
\begin{align} 
\label{eq:yukawa-result-1}
\begin{split}
&(2ik\cdot\partial_X+\delta m^2)K(k,X)\\
&\quad=\cp\Slash{k}\rho^0(k)(\nb(k^0)+\nf(k^0))\V(k,X),
\end{split}
\end{align}
where $\delta m^2\equiv \mb^2-\mf^2$.
We note that $K(k,X)$ becomes finite only when $k^2=0$ because of $\delta(k^2)$ in the right-hand side.
 We also note that $\Slash{k}K(k,X)\sim \cp^2T K(k,X)$, which is confirmed by multiplying Eq.~(\ref{eq:yukawa-result-1}) by $\Slash{k}$ from the left. 
This property makes the vertex correction term, $\Slash{k}\VC^\R(k, X)$, negligible,
which corresponds to the fact that there is no vertex correction in the analysis using the resummed perturbation theory~\cite{QED,susy-kinetic}.
Thus we get
\begin{align}
\begin{split}
\label{eq:yukawa-result} 
&(2ik\cdot\partial_X+\delta m^2)K(k,X)\\
&\quad=\cp\Slash{k}\rho^0(k)(\nb(k^0)+\nf(k^0))\varPsi(X). 
\end{split}
\end{align}
The schematic figure of $K(k,X)$ is depicted in Fig.~\ref{fig:yukawa-K}.
The solid (dashed) line with the blob stands for the resummed fermion (boson) propagator.

\begin{figure}[t]
\begin{center} 
\includegraphics[width=0.3\textwidth]{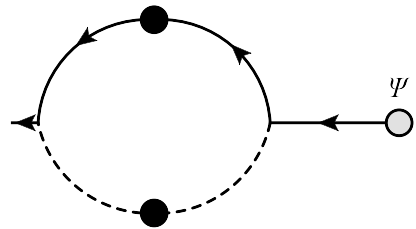}
\caption{The fermionic induced source $\etaind(X)$ in the leading order in the Yukawa model.
This diagram is obtained by connecting two ends in Fig.~\ref{fig:yukawa-K}.
By truncating $\varPsi$, 
 we obtain the diagram expressing $\varSigma$ in the resummed one-loop analysis~\cite{QED}.}
\label{fig:yukawa-induced-source}
\end{center}
\end{figure}

\subsection{Kinetic interpretation}
\label{ssc:yukawa-kinetic}
By introducing the ``{\it{off-diagonal density matrix}}'' $\varLambda_\pm(k,X)$ defined as $K(k,X)\equiv2\pi\delta(k^2)(\theta(k^0)\varLambda_+(\vk,X)+\theta(-k^0)\varLambda_-(-\vk,X))$, we arrive at the following generalized and linearized kinetic equation from Eq.~(\ref{eq:yukawa-result}):
\begin{align}
\label{eq:yukawa-kineticeq}
\begin{split}
&\left(2iv\cdot\partial_X\pm\frac{\delta m^2}{|\vk|}\right)\varLambda_\pm(\vk,X)\\
&\qquad=\cp\Slash{v}(\nb(|\vk|)+\nf(|\vk|))\varPsi(X),
\end{split}
\end{align}
where $v^\mu\equiv(1,\hat{\vk})$ and $\hat{\vk}\equiv\vk/|\vk|$.
We have to note that this equation is not a usual kinetic equation  because $\varLambda(\vk,X)$ can not be interpreted as a distribution function since it is the propagator between different particles in the fermionic background.
Nevertheless, we call this equation ``{\it{generalized kinetic equation}}''  because we can obtain the Boltzmann equation if we analyze the time-evolution of the diagonal propagator instead of the off-diagonal one~\cite{blaizot-ultrasoft}.
In fact, Eq.~(\ref{eq:yukawa-kineticeq}) has the following points similar to the Boltzmann equation: 
\begin{itemize} 
\item The particle is on-shell ($k^2\simeq 0$).
\item The equation has the common structure as the Boltzmann equation: both have the non-interacting part, the interaction part between the hard particle and the average ultrasoft field, and the interaction part among the hard particles,
which correspond to the drift term, the force term, and the collision term in the Boltzmann equation, respectively.
\end{itemize}
When $\partial_X\sim \cp T$, $\delta m^2$ is negligible,  and Eq. (\ref{eq:yukawa-kineticeq}) becomes the counterpart of the Vlasov equation~\cite{blaizot-HTL}.
Let us recapitulate the interpretations of each term in Eq. (\ref{eq:yukawa-kineticeq}) except for the $\delta m^2$ term.
The first term in the left-hand side describes the time-evolution of $\varLambda_\pm(\vk,X)$ in the free limit ($\cp=0$), so this term corresponds to the drift term in the Boltzmann equation.
On the other hand, the term in the right-hand side expresses the effect from the average fermionic field.
Hence this term corresponds to the force term in the Boltzmann equation.

$\delta m^2$ term has no counterpart in the usual Boltzmann equation, which describes the time-evolution of the diagonal propagators, $S(k,X)$ and $D(k,X)$.
Therefore we cannot obtain kinetic interpretation of that term in the usual sense.

Now let us discuss the origin of the $\delta m^2$ term.
The origin of the term is $-\{\Slash{k}, \varSigma^\R(k,X)\}+\varPi^\A(k,X)$ in Eq.~(\ref{eq:yukawa-A}).
The real parts of those terms are of order $\cp^2T^2$ while the imaginary parts are $\cp^4T\ln\cp^{-1}$, so the contribution in the leading order comes from the real parts.
The difference of the real parts of the diagonal self-energies expresses the difference of the dispersion relations of the scalar and the fermion, so we call the $\delta m^2$ term ``{\it mass difference term}.'' 
We note that the ``mass'' here is not the bare one but dynamically generated one thorough the interaction among the hard particles.

Here let us see the reason why the mass difference term does not have its counterpart in the diagonal case.
In the case of the kinetic equation on $S(k,X)$, the corresponding term to the mass difference becomes
$-\{\Slash{k}, \varSigma^\R(k,X)\}+\{\Slash{k}, \varSigma^\A(k,X)\}=-2i\im\{\Slash{k}, \varSigma^\R(k,X)\}$ 
instead of $-\{\Slash{k}, \varSigma^\R(k,X)\}+\varPi^\A(k,X)$,
so that the real part is canceled out
(See  Appendix~\ref{app:diagonal} for the detailed derivation).

We note that this term, which is pure imaginary as a result of the cancellation of the real part, is of order $\cp^4T\ln(1/\cp)$ and thus negligible since $\partial_X\sim\cp^2T$.
We also see that the terms coming from the self-energy have complicated form in the diagonal case, while they are reduced to the simple form, $-\{\Slash{k}, \varSigma^\R(k)\}+\varPi^\A(k)$ term in Eq.~(\ref{eq:yukawa-A}), in the off-diagonal case in the linear response region.
This difference comes from the following two facts:
\begin{enumerate}
\item Since we linearize the equation in terms of the deviation from the equilibrium state, the terms containing $\delta\varSigma$ and $S^0$ in Eq.~(\ref{eq:app-diagonal-kinetic}) do not have its counter parts in the off-diagonal case [$K(k,X)$ vanishes at equilibrium.]

\item The structure of the right-hand sides of Eqs.~(\ref{eq:yukawa-fermion4}) and (\ref{eq:yukawa-boson4}) after neglecting the vertex correction terms are simpler than those of Eqs.~(\ref{eq:yukawa-diagonal-1}) and (\ref{eq:yukawa-diagonal-2}).
It is because $K^<(x,y)=K^>(x,y)$. 
\end{enumerate}

\subsection{Correspondence between kinetic theory and resummed perturbation theory}
\label{ssc:yukawa-correspond}
Here, let us show the relation between our kinetic equation, Eq.~(\ref{eq:yukawa-result}), and the self-consistent equation in the resumed perturbation theory in Ref.~\cite{QED}.
For this purpose, we evaluate the retarded self-energy of the fermion, $\varSigma^R(p)$, which can be obtained from the linear response theory~\cite{blaizot-review,blaizot-HTL} in momentum space as
\begin{align}
\label{eq:yukawa-fermion-induced}
\etaind(p)=\varSigma^\R(p)\varPsi(p).
\end{align}

On the other hand, $\etaind$ can be written as
$\etaind(X)=\cp K(x,x)=\cp\int d^4k/(2\pi)^4 K(k,X)$ in our theory.
Thus, the induced source is obtained from Eq.~(\ref{eq:yukawa-result}) as
\begin{align}
\etaind(X)=\cp^2\int \frac{d^4k}{(2\pi)^4} \frac{\Slash{k}\rho^0(k)(\nb(k^0)+\nf(k^0))}{(2ik\cdot\partial_X+\delta m^2)}\varPsi(X).
\end{align}
By performing the Fourier transformation,
\begin{align}
\label{eq:fourier-p}
f(k,p)\equiv\int d^4X e^{ip\cdot X}f(k, X),
\end{align}
 we obtain
\begin{align}
\etaind(p)=\cp^2\int \frac{d^4k}{(2\pi)^4} \frac{\Slash{k}\rho^0(k)(\nb(k^0)+\nf(k^0))}{(2k\cdot p+\delta m^2)}\varPsi(p).
\label{eq:yukawa-fermion-induced2}
\end{align}
Comparing Eq.~(\ref{eq:yukawa-fermion-induced}) with Eq.~(\ref{eq:yukawa-fermion-induced2}), we obtain the self-energy,
\begin{align}
\label{eq:yukawa-selfenergy-p}
\varSigma^\R(p)=\cp^2\int \frac{d^4k}{(2\pi)^4} \frac{\Slash{k}\rho^0(k)(\nb(k^0)+\nf(k^0))}{(2k\cdot p+\delta m^2)}.
\end{align}
This expression coincides with that of the retarded fermion self-energy in Ref.~\cite{QED} except for the absence of the decay widths of the hard particles in the denominator.
As mentioned in the previous subsection, the decay widths of order $\cp^4T\ln(1/\cp)$
is neglected when the external momentum is of order  $g^2T$;
one can include them by taken into account  the imaginary part of Eq.~(\ref{eq:yukawa-result}) if one is interested in the decay width.

The diagrammatic representation of the fermion retarded self-energy in our formalism is the same as that in the resummed perturbation theory~\cite{QED}, which is explained as follows:
The off-diagonal density matrix $\varLambda_\pm(\vk,X)$, which follows the generalized kinetic equation, is represented by Fig.~\ref{fig:yukawa-K}.
$\etaind(X)$, shown in Fig.~\ref{fig:yukawa-induced-source}, is diagrammatically obtained by connecting the ends of fermion and boson propagators in the right-hand side of Fig.~\ref{fig:yukawa-K}.
This diagram is the resummed one-loop diagram appeared in~\cite{QED} itself except for the fermion average field $\varPsi$.

\section{Kinetic equation in QED}
\label{sec:QED}
First we introduce the background field method, which is useful to construct the equations for the average fields and the Kadanoff-Baym equation in a gauge-covariant form.
Next, we derive the generalized kinetic equation in the linear response regime adopting the Coulomb gauge fixing, in which the transversality of the free photon propagator simplifies the analysis.
After the derivation, we show the equivalence between the generalized kinetic equation and the resummed perturbation theory~\cite{lebedev,QED}, and discuss the interpretation of the terms in the kinetic equation.
We also check that the Ward-Takahashi identity for the self-energy, consequence of the $U(1)$ gauge symmetry, is satisfied in our formalism.
Finally we discuss how to compute the higher-point-vertex function whose external momenta are all ultrasoft, and make an order estimate of it at the weak coupling regime.

\subsection{Background field gauge method}

In the derivation of the average field equation and the Kadanoff-baym equation in QED, 
it is convenient to formulate them in a covariant form under gauge transformations.
For this purpose, we employ the background field gauge method~\cite{blaizot-review,background-field-gauge}.
In this method, the following generating functional is employed:
\begin{align}
\begin{split}
\tilde{Z}[j,\eta,\overline{\eta}; A,\varPsi,\overline{\varPsi}]=\int [{\cal D}{a}][{\cal D}\overline{\psi}][{\cal D}\psi]e^{ iS},
\label{eq:generatingFunctionalQED}
\end{split}
\end{align}
with
\begin{equation}
\begin{split}
S&= \int_C d^4x 
\Bigl[  {\cal L}[A_\mu+a_\mu, \varPsi+\psi, \overline{\varPsi}+\overline{\psi}]+{\cal {L}}_{\mathrm {GF}}\\
&\quad-(j^\mu a_\mu+\overline{\psi}\eta+\overline{\eta}\psi)\Bigr] ,
\end{split}
\end{equation}
where we dropped the ghost  term, which is not coupled with the other fields.
$A^\mu$ and $a^\mu$ are vector fields, $\varPsi$ ($\overline{\varPsi}$) and $\psi$ ($\overline{\psi}$) are (anti-) spinor fields, and $j^\mu$ is the external current, respectively.
The Lagrangian of QED has the form,
\begin{align}
{\cal L}[a, \psi, \overline{\psi}]=-\frac{1}{4}F^{\mu\nu}[a]F_{\mu\nu}[a] +i\overline{\psi}\Slash{D}[a]\psi,
\end{align}
where $F_{\mu\nu}[a]\equiv \partial_\mu a_\nu-\partial_\nu a_\mu $ is the field strength and $D_\mu[a]\equiv \partial_\mu +i\cp a_\mu$ is the covariant derivative.
We have used $\cp$ as a coupling constant of QED instead of the standard notation $e$ to make it clear that the same order counting appears as that in the Yukawa model.
In the background field method, the fields in the Lagrangian are decomposed to the classical field, identified as the average fields later, and fluctuations in Eq.~(\ref{eq:generatingFunctionalQED}). 
The external sources are chosen to be coupled to $a^\mu$, $\psi$, and $\overline{\psi}$, but not to $A^\mu$, $\varPsi$, and $\overline{\varPsi}$. 
We impose the following conditions:
\begin{align}
\label{eq:QED-condition}
\langle a_\mu\rangle=\langle\psi\rangle=\langle\overline{\psi}\rangle=0,
\end{align}
which implies that $A_\mu$ and $\varPsi$ ($\overline{\varPsi}$) can be interpreted as the average parts of the photon and (anti) electron field, respectively, and $-\ln \tilde{Z}$ coincides with the effective action~\cite{blaizot-review,background-field-gauge}.

In the background gauge field method, the gauge-fixing term is chosen to be a functional of $a_\mu$ such as
\begin{align}
{\cal {L}}_{\mathrm {GF}}= -\lambda\frac{(\GF[a])^2}{2},
\end{align}
where $\GF[a]$ is the gauge-fixing function and $\lambda$ is the gauge-fixing parameter.
Although $a_\mu$ is fixed by the gauge fixing term,
the generating functional, Eq.~(\ref{eq:generatingFunctionalQED}),  is invariant under the background field gauge transformations defined by
\begin{align}
\label{eq:qed-gauge-trans}
\begin{split}
\varPsi(x)&\rightarrow h(x)\varPsi(x),~~ 
A_\mu(x)\rightarrow A_\mu(x)-\frac{i}{\cp}h(x)\partial_\mu h^\dagger(x), ~~\\
\psi(x)&\rightarrow h(x)\psi(x) ,~~ 
\overline{\psi}(x)\rightarrow  \overline{\psi}(x)h^\dagger(x),~~
a_\mu(x)\rightarrow a_\mu(x),\\
\eta(x)&\rightarrow h(x)\eta(x) ,~~
\overline{\eta}(x)\rightarrow \overline{\eta}(x)h^\dagger(x),~~
j^\mu(x)\rightarrow j^\mu(x),
\end{split}
\end{align}
where $h(x)\equiv \exp[{i\theta(x)}]$.

Since the fluctuations covariantly transform under Eq.~(\ref{eq:qed-gauge-trans}), the propagators also covariantly transform as
\begin{align}
D_{\mu\nu}(x,y)&\equiv  \langle\Tc a_\mu(x)a_\nu(y)\rangle_c\to D_{\mu\nu}(x,y),\\
S(x,y)&
\to h(x)S(x,y) h^\dagger(y),\\
K_\mu(x,y) &\equiv \langle\Tc\psi(x)a_\mu(y)\rangle_c \to  h(x)K_\mu(x,y).
\end{align}

Also, the Wigner transformed off-diagonal propagator in the leading order of $\cp$ is covariant, which can be confirmed by performing the gradient expansion~\cite{blaizot-review}:
\begin{align}
\label{eq:QED-K-transform}
K_\mu(k,X)\rightarrow h(X) K_\mu(k,X),
\end{align}
which implies that the Kadanoff-Baym equation covariantly transforms  with respect to the background gauge transformations as will be seen in the next subsection.

We note that, apart from the covariance with respect to the background field gauge transformation, the gauge-fixing dependence, which will be confirmed only in the Coulomb gauge and temporal gauge in this paper, should be analyzed. 

\subsection{Derivation of the kinetic equation} 
\label{ssc:derivation}
We work in the Coulomb gauge-fixing condition because this gauge fixing makes the analysis simple owing to the transversality of the free photon propagator.
The gauge-fixing condition is $\GF[a]=\partial_i a^i$ and $\lambda\rightarrow \infty$,
which constrain  the off-diagonal propagator as
\begin{align}
\label{eq:qed-coulomb-condition}
\partial^i_y K_i (x,y)=0.
\end{align}

The equations of motion for the average fields are given by
\begin{align}
\label{eq:qed-meanfield-fermion}
i\Slash{D}_x[A]\varPsi(x)&=\eta(x)+\etaind(x),\\ 
\notag
\partial^2A^\mu(x)-\partial^\mu\partial^\nu A_\nu(x)&-\cp\overline{\varPsi}(x)\gamma^\mu\varPsi(x)\\
\label{eq:qed-meanfield-boson}
&=j^\mu(x)+\jind^\mu(x).
\end{align}
Here we have used Eq.~(\ref{eq:QED-condition}), and the induced fermionic source and the induced current are defined as
\begin{align}
\etaind(x)&\equiv \cp\langle\Slash{a}(x)\psi(x)\rangle_c =\cp\gamma^\mu K_\mu(x,x) ,\\
\jind^\mu(x)&\equiv \cp\langle\overline{\psi}(x)\gamma^\mu\psi(x)\rangle_c =\cp \text{Tr}(\gamma^\mu S^<(x,x)),
\end{align}
which transform  $\etaind(x)\rightarrow h(x)\etaind(x)$ and $\jind^\mu(x)\rightarrow \jind^\mu(x)$ under the background gauge transformations.
Therefore, Eqs.~(\ref{eq:qed-meanfield-fermion}) and (\ref{eq:qed-meanfield-boson}) transform covariantly with respect to the background gauge transformation.

The equations for the propagators are given by 
\begin{align}
\notag
&\Slash{D}_xK^\mu(x,y)+i\cp\gamma_\nu D^{\nu\mu}(x,y)\varPsi(x)\\
\label{eq:qed-fermion}
&\quad=-i\int^\infty_{-\infty} d^4z (\varSigma^\R(x,z)K^\mu (z,y)\\
\notag
&\qquad+\VC^\R_{\nu}(x,z) D^{< \nu\mu}(z,y)) , \\ 
\notag
&(\partial^2 g^{\mu \nu}-\partial^\mu\partial^\nu)_yK_\nu(x,y)+\cp S^<(x,y)\gamma^\mu\varPsi(y)\\
\label{eq:qed-boson}
&\quad=\int^\infty_{-\infty} d^4z (\varPi^{\A\mu\nu}(z,y)K_\nu(x,z)-S^<(x,z)\VC^{\R\mu}(z,y)),
\end{align}
where  $\varPi_{\mu\nu}(x,y)$ and $\VC_\mu(x,y)$ are 
  the photon and the off-diagonal self-energies, respectively.  
Here we set $x^0\in C^+$ and $y^0\in C^-$.

The Wigner transformed equations read 
\begin{align}
\notag
&\left(-i\Slash{k}+\frac{\Slash{\partial}_X}{2}+i\cp\Slash{A}(X)\right)K^\mu(k,X)\notag\\
\label{eq:qed-wigner-fermion}
&\qquad+i\cp\gamma_\nu D^{<\nu\mu}(k,X)\varPsi(X) \notag\\
&\quad=-i(\varSigma^\R(k,X)K^\mu (k,X)+\VC^\R_{\nu}(k,X) D^{< \nu\mu}(k,X)), \\
\notag
&(-k^2+ik\cdot\partial_X)K^{\mu}(k,X)+k^\mu k^0K_0(k,X) \notag\\
&\qquad+\cp S^<(k,X)\gamma^\mu\varPsi(X) \notag\\
\label{eq:qed-coulomb-wigner-boson}
&\quad=\varPi^{\A\mu\nu}(k,X)K_\nu(k,X)-S^<(k,X)\VC^{\R \mu}(k,X),
\end{align}
where Eq.~(\ref{eq:qed-coulomb-condition}), the estimation $K^0\sim\cp^2 K^i$, shown in Appendix~\ref{app:K0}, and the gradient expansion have been used.
Multiplying Eq.~(\ref{eq:qed-wigner-fermion}) by $[-i\Slash{k}+\Slash{\partial}_X/2+i\cp\Slash{A}(X)+i\varSigma^\R(k,X)]$ and Eq.~(\ref{eq:qed-coulomb-wigner-boson}) by $P^{T}_{\mu i}(k)$ defined below, and subtracting the latter from the former, we obtain
\begin{align}
\label{eq:qed-coulomb-K}
\begin{split}
&(-2ik\cdot\partial_X+2\cp k\cdot A(X)+\{\Slash{k},\varSigma^\R(k,X)\})K^i(k,X)\\
&\qquad+P^{T }_{\alpha i}(k)\varPi^{\A \alpha\nu}(k,X) K_\nu(k,X)\\
&\quad=
-\cp(\Slash{k}D^{<  \nu i} (k,X)+P^{T  \nu i} (k)S^<(k,X))\V_\nu(k,X),
\end{split}
\end{align}
where we have introduced the projection operator into the transverse component, $P^T_{ \mu\nu}(k)\equiv g_{\mu i}g_{\nu j}(\delta_{ij}-\hat{k}_i\hat{k}_j)$
, and ${\cp}\V^\mu(k,X)\equiv \cp\gamma^\mu\varPsi(X)+\VC^{\R\mu}(k,X)$, and used $P^{T}_{\mu i}(k)K^\mu(k,X)=-K_i(k,X)=K^i(k,X)$. 
One can show that the background fields and the coupling dependences in the diagonal propagators are weak\footnote{It can be shown by evaluating $\delta S$ and $\delta D^{\mu\nu}$ in the diagonal kinetic equation: $\delta S\sim \cp A^\mu$ from Eq.~(\ref{eq:app-diagonal-kinetic-QED}).
Similarly, $\delta D^{\mu\nu}\sim \delta S$.
These order estimates imply that $A^\mu$ contribution from $S$ and $D^{\mu\nu}$ in the right-hand side of Eq.~(\ref{eq:qed-coulomb-K}) is much smaller than that from the left-hand side of that equation, by using the order estimate $K^\mu\sim \cp^{-1}\Psi$.
},
so that we can replace the electron and the photon propagator, which is given as follows, by that in the free limit at equilibrium:
\begin{align}
D^{0<}_{\mu\nu}(k)=&\rho^0(k)\nb(k^0)P^T_{\mu\nu}(k).
\end{align}
The diagonal self-energies at on-shell in the leading order are given by 
\begin{align}
\{\Slash{k}, \varSigma^{\R ({\mathrm {eq}})}(k) \}&=\me^2-2i\zetae k^0,\\
P^{T }_{\alpha i}(k)\varPi^{\A  ({\mathrm {eq}}) \alpha\nu}(k)&= -\mg^2 P^{T \nu}_i(k), 
\end{align} 
where $\me\equiv \cp T/2$ and $\mg\equiv \cp T/\sqrt{6}$ are the asymptotic thermal masses of the electron and the photon in the leading order.
$\zetae\sim \cp^2 T\ln (1/\cp)$ is the damping rate of the hard electron~\cite{electron-damping}, which is much larger than the case of the Yukawa model.
Unlike in the case of Yukawa model, the damping rate of the electron $\zetae$ cannot be neglected because $\zetae k^0\sim \me^2$, while 
 the photon damping rate of order $\cp^4 T\ln(1/\cp)$ can be neglected~\cite{photon-damping}.
We note that the longitudinal part of the photon self-energy does not contribute because the projection operator $P^{T i}_{\alpha}(k)$ is multiplied.

The off-diagonal self-energy in the leading order has the following expression, which has a similar form to that in the Yukawa model:
\begin{align}
\label{eq:qed-vertex-correction}
\VC^{\R\mu}(k,X)= -\cp^2\int\frac{d^4 k'}{(2\pi)^4}\gamma^\nu S^{0\R}(k+ k')\gamma^\mu K_\nu(k',X) .
\end{align}
The off-diagonal self-energy  in $\V^\mu(k,X)$ has to be retained in the case of QED unlike the Yukawa model since there is no special suppression mechanism.

By substituting these expressions in Eq.~(\ref{eq:qed-coulomb-K}), we obtain
\begin{align}
\label{eq:qed-coulomb-result-K}
\begin{split}
&(-2ik\cdot D_X-2i\zetae k^0-\delta m^2)K^i(k,X)\\
&\quad=-\cp\Slash{k} P^{T \nu i}(k) \rho^0(k)(\nb(k^0) +\nf(k^0))\\
&\qquad\times\left(\gamma_\nu\varPsi(X)
+ \cp\int\frac{d^4 k'}{(2\pi)^4}\frac{k^\alpha \gamma_\nu+\gamma^\alpha k'_\nu}{k\cdot k'} K_\alpha(k',X)\right).
\end{split}
\end{align}
Here we have used $P^T_{\mu\alpha}(k) P^{T \alpha}_{\nu}(k)=-P^T_{\mu\nu}(k) $ and introduced $\delta m^2\equiv \mg^2-\me^2$.
In this gauge-fixing condition, it is apparent that only the transverse component of the thermal photon contributes to $K^i(k,X)$ because of the projection operator $P^{T \nu i}(k)$ appearing in the right-hand side of Eq.~(\ref{eq:qed-coulomb-result-K}). 
We note that this equation transforms covariantly with respect to the background gauge transformation from Eq.~(\ref{eq:QED-K-transform}).

\begin{figure*}[t]
\begin{center}
\includegraphics[width=0.8\textwidth]{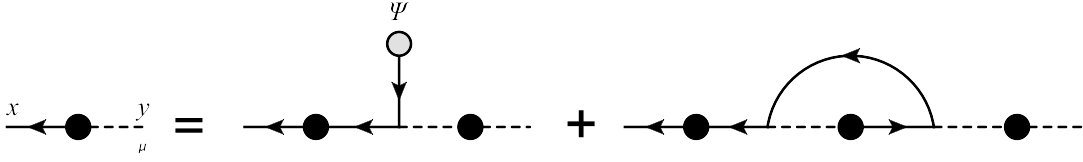}
\caption{The diagrammatic representation of self-consistent equation for $K^\mu(x,y)$ in the leading order.
For simplicity, $A^\mu$ is not drawn.
}
\label{fig:QED-K}
\end{center}
\end{figure*}

From Eq.~(\ref{eq:qed-coulomb-result-K}), we write $K^i(k,X)$ in terms of the off-diagonal self-energy for later use:
\begin{align}
\label{eq:qed-coulomb-K-vertex}
&K^i(k,X)
={\cp}\frac{\Slash{k}P^{T i}_{\nu}(k)  \rho^0(k)(\nb(k^0) +\nf(k^0))}{2ik\cdot D_X +2i\zetae k^0+\delta m^2}\V^\nu(k,X).
\end{align}
The diagrammatic representation of this equation is shown in Fig.~\ref{fig:QED-K}.

\subsection{Kinetic interpretation}

Next, we derive  the linearized kinetic equation.
Multiplying Eq.~(\ref{eq:qed-coulomb-result-K}) by $\gamma_i$ from the left, we obtain
\begin{align}
\label{eq:QED-coulomb-result-Kslash}
\begin{split}
&(-2ik\cdot D_X-2i\zetae k^0-\delta m^2)\Slash{K}(k,X)\\
&\quad=-\cp 2\Slash{k} \rho^0(k)(\nb(k^0) +\nf(k^0))\varPsi(X)\\
&\qquad+\gamma_i\Slash{k}P^{T i}_{\nu}(k)  \rho^0(k)(\nb(k^0) +\nf(k^0))\\
&\qquad\times\cp^2\int\frac{d^4 k'}{(2\pi)^4}\frac{k^\alpha\gamma^\nu+k'^\nu\gamma^\alpha}{k\cdot k'}K_\alpha(k',X).
\end{split}
\end{align}
We  decompose $\varLambda^\mu_\pm(\vk,X)$ into positive and negative energy terms as $K^\mu(k,X)\equiv2\pi\delta(k^2)[\theta(k^0)\varLambda^\mu_+(\vk,X)+\theta(-k^0)\varLambda^\mu_-(-\vk,X)]$, so that we arrive at the kinetic equation from Eq.~(\ref{eq:QED-coulomb-result-Kslash}):
\begin{align}
\label{eq:QED-kineticeq}
\begin{split}
&\left(2iv\cdot\partial_X\pm\frac{\delta m^2}{|\vk|}+2i\zetae\right)\Slash{\varLambda}_{\pm}(\vk, X)\\
&\quad=2\cp \Slash{v}[\nb(|\vk|)+\nf(|\vk|)]\varPsi(X)\\
&\qquad-\cp^2\gamma_{i}\Slash{v}[\nf(|\vk|)+\nb(|\vk|)]P^{\nu i }_{\mathrm T}(v)\\
&\qquad\times\sum_{s=\pm}\int\frac{d^3\vk '}{(2\pi)^3}\frac{1}{2|\vk '|}\frac{s|\vk|v^\alpha\gamma_\nu\pm |\vk'|v_{ \nu}'\gamma^\alpha}{|\vk| |\vk '|v\cdot v'}\varLambda_{s \alpha}(\vk',X),
\end{split} 
\end{align}
where we have introduced $v^{\prime\mu}\equiv (1, \hat{\vk} ')$.
There are two terms that do not appear in the Yukawa model analyzed in Sec.~\ref{sec:yukawa}.
One is the last term in the right-hand side.
Because the bare vertex term, which contains $\varPsi(X)$ and is in the right-hand side, is interpreted as the counterpart of the force term in the diagonal case~\cite{blaizot-HTL}, that vertex correction term acts like ``{\it  the correction to the force term},'' at least in the linear response regime.  
Note that this term mixes the positive and negative energy modes, unlike the case of the Yukawa model.

The other is the third term in the left-hand side.
 This term has a similar form to the collision term in the relaxation time approximation of the diagonal case, i.e., pure imaginary constant ($2i\zetae$) times $\varLambda_\pm(\vk,X)$.
For this reason, we call this term ``{\it collision term}.''
We note that this term is negligible in the case of the Yukawa model as shown in the previous section.

In the diagonal case~\cite{blaizot-ultrasoft}, the collision term contains momentum integral for the diagonal density matrix\footnote{The integral comes from the terms containing $\delta\varSigma$ in Eq.~(\ref{eq:app-diagonal-kinetic-QED})}.
In contrast, in the off-diagonal case, such term does not survive in the linearized equation.
As a result, the collision term in the off-diagonal kinetic equation has a similar form to that in the relaxation time approximation.

We emphasize that the off-diagonal self-energy is not negligible in the off-diagonal kinetic equation, while negligible in the diagonal one because we neglect the higher order terms in $\varPsi$. 
This fact makes the correction to the force term, which is absent in the diagonal case, appears in Eq.~(\ref{eq:QED-kineticeq}).

 As we discussed in Sec.~\ref{ssc:yukawa-kinetic}, both of the usual Boltzmann equation and our generalized and linearized kinetic equation are composed of the non-interacting part, the interaction part between the hard particle and the average ultrasoft field, and the interaction part among the hard particles.
Which part is the counterpart of the mass difference term, the collision term, and the correction to the force term?
Because the mass difference and the collision term come from the self-energies at equilibrium, they correspond to the interaction part among the hard particle.
The correction to the force term is a part of the interaction part between the hard particle and the average ultrasoft field.


\begin{table}[t]
\begin{center}
\caption{The correspondence between the resummed perturbation theory and the generalized and linearized kinetic equation.}
\begin{tabular}{c|c}
\hline
Diagrammatic method & kinetic equation  \\ \hline \hline
thermal mass difference & mass difference term \\
decay width& collision term \\
ladder diagrams&  correction to force term \\
\hline
\end{tabular}
\label{tab:}
\end{center}
\end{table}

\subsection{Correspondence between kinetic theory and resummed perturbation theory}
Here, let us show the equivalence between  Eq.~(\ref{eq:qed-coulomb-result-K}) and the self-consistent equation in the resummed perturbation theory~\cite{QED,lebedev}.
To this end, we rewrite Eq.~(\ref{eq:qed-vertex-correction}) in terms of the off-diagonal self-energy using Eq.~(\ref{eq:qed-coulomb-K-vertex}):
\begin{align}
\begin{split}
\V^\mu(k,X)&=\gamma^\mu\varPsi(X)
 -\cp^2\int\frac{d^4 k'}{(2\pi)^4}\gamma^\nu S^{0\R}(k+ k')\gamma^\mu\\
 &\quad\times \frac{\Slash{k}'P^{T }_{\alpha\nu}(k')  \rho^0(k')[\nb(k'^{0}) +\nf(k'^{0})]}{2ik'\cdot \partial_X +2i\zetae k'^{0}+\delta m^2}\\
 &\quad \times\V^\alpha(k',X) .
\end{split}
\end{align}
Here we set $A^\mu=0$.
By performing the Fourier transformation, Eq.~(\ref{eq:fourier-p}), we get
\begin{align}
\label{eq:QED-BS}
\begin{split}
\varGamma^\mu(k,p)&=\gamma^\mu
 -\cp^2\int\frac{d^4 k'}{(2\pi)^4}\gamma^\nu S^{0\R}(k+ k')\gamma^\mu\\
 &\!\!\!\quad\times \!\frac{\Slash{k}'\! P^{T }_{\alpha\nu}(k')  \rho^0(k')[\nb(k'^{0}) +\nf(k'^{0})]\varGamma^\alpha(k',p)}{2k'\cdot p +2i\zetae k'^{0}+\delta m^2} ,
\end{split}
\end{align}
where $\V^{\mu}(k,p)\equiv \varGamma^\mu(k,p)\varPsi(p)$.
We note that $\varGamma^\mu(k,p)$ is the vertex function introduced in \cite{QED,lebedev} whose momenta are hard and ultrasoft.
Equation~(\ref{eq:QED-BS}) is none other than the integral equation appearing in~\cite{lebedev, QED}.

The retarded fermion self-energy is also written in terms of the vertex function:
from Eq.~(\ref{eq:qed-coulomb-K-vertex}), we arrive at
\begin{align}
\begin{split}
\varSigma^\R(p)&= \cp\int \frac{d^4k}{(2\pi)^4} \frac{\delta\Slash{K}(k,p)}{\delta\varPsi(p)}\\
&=\cp^2\int \frac{d^4k}{(2\pi)^4}\frac{\gamma_i\Slash{k}P^{T i}_{\nu}(k)  \rho^0(k)[\nb(k^0) +\nf(k^0)]}{2k\cdot p +2i\zetae k^0 +\delta m^2}\\
&\quad\times \varGamma^\nu(k,p).
\end{split}
\end{align}
This expression equals to that of the fermion retarded self-energy in the resummed perturbation theory\footnote{Equation~(\ref{eq:QED-BS}) is analytically solved for the energy region $|p_0+i\zetae |\ll \cp^2T$~\cite{QED}.}~\cite{lebedev,QED}.
Thus we see that Eq.~(\ref{eq:qed-coulomb-result-K}), derived in the non-equilibrium state in a linear response regime from the Kadanoff-Baym equation, is equivalent to 
the self-consistent equation in the resummed perturbation theory, which is constructed in the thermal equilibrium state.

Here we discuss the correspondence between each scheme in the resummed perturbation theory~\cite{lebedev,QED} and each term in the kinetic equation.
As in the Yukawa model, the resummation of the thermal mass difference in the resummed perturbation theory corresponds to the mass difference term in the kinetic equation.
The decay width corresponds to the collision term since both of them contains the damping rate of the hard electron, $\zetae$.
The ladder summation in the resummed perturbation theory~\cite{QED,lebedev} is caused by the correction to the force term in the kinetic equation.
Thus, the ladder summation corresponds to the correction to the force term.
These interpretations are summarized in the Table.~\ref{tab:}.

\subsection{Higher point functions} 

The fermionic induced source $\etaind$ generates the higher point function, not only the fermion self-energy.
In this subsection, we derive the self-consistent equation determining the $n$-point function
whose external lines consist of two fermions ($\varPsi$) and $(n-2)$ bosons ($A^\mu$) with ultrasoft external momenta, and make an order estimate of the quantity.
For example in the case of $n=3$, we obtain the correction to the bare three-point function, $\cp\gamma_\mu \delta^{(4)}(p-q-r)$, from $\etaind$~\cite{blaizot-HTL,blaizot-review}:
\begin{align}
\label{eq:QED-ultrasoft-vertex}
\begin{split}
\delta^{(4)}&(p-q-r)\cp\delta\varGamma^\mu(p,-q,-r)\\
&\equiv\left.\frac{\delta^2 \etaind(p)}{\delta\varPsi(q) \delta A_\mu(r)}\right|_{A=0}\\
&= \cp \int\frac{d^4k'}{(2\pi)^4}\frac{\delta^2 }{\delta\varPsi(q) \delta A_\mu(r)}\Slash{\delta K}(k,p).
\end{split}
\end{align}
Here we have expanded $K^\mu(k,X)$ around $A^\mu=0$: 
\begin{align} 
K^\mu(k,X)=K^\mu(k,X)_{A=0}+\delta K^\mu(k,X) +O(A^2\varPsi),
\end{align} 
where $K^\mu(k,X)_{A=0}$ contains one $\varPsi$ while $\delta K^\mu(k,X)$ contains one $\varPsi$ and one $A^\mu$.

$\delta K^\mu(k,X)$ can be obtained by the following way.
Collecting terms that contain one $A^\mu$ in Eq.~(\ref{eq:qed-coulomb-result-K}), we obtain
\begin{align}
\begin{split}
&(-2ik\cdot \partial_X-2i\zetae k^0-\delta m^2)\delta K^i(k,X)\\
&\qquad+2\cp k\cdot A(X) K^i(k,X)_{A=0}\\
&\quad=-\cp^2\Slash{k} P^{T \nu i}(k) \rho^0(k)[\nb(k^0) +\nf(k^0)]\\
&\qquad\times \int\frac{d^4 k'}{(2\pi)^4}\frac{k^\alpha \gamma_\nu+\gamma^\alpha k'_\nu}{k\cdot k'} \delta K_\alpha(k',X). \label{eq:deltaKi}
\end{split}
\end{align}
Since $K^\mu(k,X)_{A=0}$ is determined by setting $A^\mu=0$ in Eq.~(\ref{eq:qed-coulomb-result-K}), this equation is closed and $\delta K^\mu(k,X)$ can be determined.

Let us estimate  the order of $\delta K^\mu(k,X)$.
From Eq.~(\ref{eq:qed-coulomb-result-K}) and $K^\alpha_{A=0}\sim \cp^{-1}T^{-3}\varPsi $, we 
find $\delta K^\mu\sim \cp^{-1}T^{-1}A^\mu K^\alpha_{A=0}\sim \cp^{-2}T^{-4}\varPsi A^\alpha$.
Therefore, the vertex correction is estimated as $\cp\delta\varGamma^\mu\sim \cp^{-1}$, which is much larger than the bare vertex, $\cp\gamma^\mu\sim\cp$.
Similar order estimate for the $n$-point function with $n>3$ can be done 
with the same procedure; as a result, we find that the order of the $n$-point-vertex function is $\cp^{2-n}$ .

\subsection{Ward-Takahashi identity}

Let us see that the off-diagonal self-energy given in Eq.~(\ref{eq:qed-vertex-correction}) satisfies the Ward-Takahashi (WT) identity.
From Eq.~(\ref{eq:qed-vertex-correction}), we get
\begin{align}
\label{eq:qed-WT}
\begin{split}
 &k_\mu\VC^{\R \mu}(k,X)\\
&= \cp^2\int\frac{d^4 k'}{(2\pi)^4}\gamma^\nu \frac{\Slash{k}+\Slash{k}'}{(k+k')^2}(\Slash{k}+\Slash{k}'-\Slash{k}')K_\nu(k',X) \\
&= \cp^2\int\frac{d^4 k'}{(2\pi)^4}\Slash{K}(k',X)=\cp\etaind(X).
\end{split}
\end{align}
Here we have used $\Slash{k}'K_\nu(k',X)=0$, which can be confirmed by multiplying Eq.~(\ref{eq:qed-coulomb-result-K}) by $\Slash{k}$ from the left.
This equation generates the WT identity derived in Ref.~\cite{QED} by setting $A^\mu=0$.
The WT identity implies that the vertex correction is not negligible because the identity relates the vertex correction to the fermion self-energy, which is much larger than the inverse of the fermion propagator with an ultrasoft momentum.
In the Yukawa model, the WT identity associated with gauge symmetries is absent, so the smallness of the vertex correction is not in contradiction with any identity.

Equation~(\ref{eq:qed-WT}) can be derived from the conservation law of the induced current, 
\begin{align}
-i\cp(\overline{\eta}_{\text {ind}}\varPsi-\overline{\varPsi}\etaind)(x)
=\partial_\mu \jind^\mu(x).
\label{eq:conservationLaw}
\end{align}
By differentiating Eq.~(\ref{eq:conservationLaw}) with respect to $ \overline{\varPsi}(y)$, we obtain
\begin{align}
\label{eq:QED-WT-pre}
\begin{split}
&-\cp\left(\frac{\delta{\overline{\eta}_{\text {ind}}}(x)}{\delta\overline{\varPsi}(y)}\varPsi(x)-\delta^{C}(x^0-y^0)\delta^{(3)}(\vx-\vy)\etaind(x)\right)\\
&\quad=\partial^x_\mu \VC^\mu(y,x).
\end{split}
\end{align}
Here $\delta^{C}(x-y)$ is the delta function defined along the contour $C$.
By multiplying this equation by $\int d^4s \exp({ik\cdot s})$ and taking only the leading-order terms, we find 
\begin{align}
\label{eq:QED-WT2}
i\cp\etaind(X)
=-i k_{\mu}\VC^{\R \mu}(-k,X). 
\end{align} 
This is nothing but  Eq.~(\ref{eq:qed-WT}). The detail of the derivation of Eq.~(\ref{eq:QED-WT2}) is shown in the Appendix \ref{app:analyticity}. 

\section{Summary and Concluding Remarks}
\label{sec:summary}

We derived the novel generalized and linearized kinetic equation with an ultrasoft momentum from the Kadanoff-Baym equation in the Yukawa model and QED.
Our kinetic equation is equivalent to the self-consistent equation in the resummed perturbation theory~\cite{QED,lebedev} used in the analysis of the fermion propagator.
This derivation helps us to establish the foundation of the resummed perturbation theory. 
We gave the kinetic interpretation of the resummation scheme (Table.~\ref{tab:}).
Furthermore, we made an order estimate of the higher-point function with ultrasoft external momenta, and obtained the equation determining the vertex correction in QED.

In our analysis, the difference of the masses and the vertex correction, which reflects the fact that we analyze a process that changes the type of particle, plays an important role. 
This fact suggests that these quantities may play an important role in the analysis of other ultrasoft quantity such as the flavor diffusion constant in QCD, where the masses of $u$, $d$ and $s$ are different. It is natural to ask whether such kinetic equation is derived in QCD because the investigation of the QCD at high temperature is quite interesting since it is relevant to analysis of quark-gluon plasma~\cite{QGP}, which is realized in heavy ion collision experiments. 
We hope to report the analysis in QCD elsewhere~\cite{QCD}.

\section*{Acknowledgement}

This work was supported by the Grant-in-Aid for the Global COE Program ``The Next Generation of Physics, Spun from Universality and Emergence'' from the Ministry of Education, Culture, Sports, Science and Technology (MEXT) of Japan.
Y.H. is supported by a Grant-in-Aid for Scientific Research (No.23340067) from
MEXT.
\appendix
\section{ANALYTIC STRUCTURE OF INDUCED TERMS}
\label{app:analyticity}

In this Appendix, we derive the right-hand sides of Eqs.~(\ref{eq:yukawa-fermion3}), (\ref{eq:yukawa-boson3}), and (\ref{eq:QED-WT2}).
First let us derive the right-hand sides of Eqs.~(\ref{eq:yukawa-fermion3}) and (\ref{eq:yukawa-boson3}):
\begin{align}
\label{eq:app-ana-1}
\begin{split}
&\int_C d^4z \varSigma(x,z) K(z,y)\\
&\quad=\int^{x^0}_{t^0} d^4z \varSigma^>(x,z) K(z,y) 
- \int^{y^0}_{x^0} d^4z \varSigma^<(x,z) K(z,y)\\
&\qquad-\int^{t^0-i\beta}_{y^0} d^4z \varSigma^<(x,z) K(z,y)\\
&\quad=\int^{x^0}_{t^0} d^4z (\varSigma^>(x,z)+\varSigma^<(x,z)) K(z,y) \\
&\qquad-\int^{t^0-i\beta}_{t^0} d^4z \varSigma^<(x,z) K(z,y)\\
&\quad\simeq-i\int^{\infty}_{-\infty} d^4z \varSigma^{R}(x,z) K(z,y).
\end{split}
\end{align}
In the last line we have taken $t^0\rightarrow-\infty$.
We used the fact that the term integrated on $C^0$ becomes negligible in this limit~\cite{blaizot-review}.
In the same way, we get
\begin{align}
\label{eq:app-ana-2}
\int_C d^4z \varPi(y,z) K(x,z)\simeq&-i\int^{\infty}_{-\infty} d^4z  \varPi^{A}(z,y)K(x,z).
\end{align}

Next, we evaluate the off-diagonal self-energy term.
The vertex correction term of Eq.~(\ref{eq:yukawa-fermion2}) becomes
\begin{align}
\label{eq:app-ana-3}
\begin{split}
&\int_C d^4z  \VC(x,z)D(z,y)\\
&\quad=\int^{x^0}_{t^0} d^4z \VC^>(x,z) D^<(z,y) \\
&\qquad+ \int^{y^0}_{x^0} d^4z \VC^<(x,z) D^<(z,y)\\
&\qquad+\int^{t^0-i\beta}_{y^0} d^4z\VC^<(x,z) D^>(z,y)\\
&\quad= \int^{x^0}_{t^0} d^4z (\VC^>(x,z)-\VC^<(x,z)) D^<(z,y) \\
&\qquad+ \int^{y^0}_{t^0} d^4z \VC^<(x,z)  (D^<(z,y)-D^>(z,y))\\
&\qquad-\int^{t^0-i\beta}_{t^0} d^4z \VC^<(x,z) D^>(z,y)\\
&\quad\simeq-i\int^{\infty}_{-\infty} d^4z (\VC^{R}(x,z) D^<(z,y)\\
&\qquad+\VC^<(x,z)D^\A(z,y)),
\end{split}
\end{align}
where the advanced boson propagator is $D^\A(z,y)\equiv -i\theta(y^0-z^0)[D^>(z,y)-D^<(z,y)]$.
Here we stop and discuss the structure of the off-diagonal self-energy in the leading order in the $(k,X)$ space.

The off-diagonal self-energy in the leading order and linear response regime is given by 
\begin{align}
\VC(x,y)=\cp^2 S^{0}(x,y)K(y,x),
\end{align}
where $S^{0}(x,y)$ is the free fermion propagator at equilibrium.
Thus, the components of $\VC$ are given by
\begin{align}
\VC^{\gtrless}(x,y)=
\pm \cp^2 S^{0 \gtrless}(x,y)K(y,x).
\end{align}
By performing the Wigner transformation, we get
\begin{align}
\label{eq:app-analyticity-vertex-correction-k}
\VC^\gtrless(k,X)=\pm \cp^2\int\frac{d^4k'}{(2\pi)^4}S^{0 \gtrless}(k+k')K(k', X),
\end{align} 
with $S^{0 >}(k)\equiv \Slash{k}\rho^0(k)(1-\nf(k^0))$.
We see that since $K(k', X)$ contains $\delta(k'^2)$ and we focus on the on-shell case $k^2\simeq 0$, which are confirmed from the analysis in Sec.~\ref{ssc:yukawa-derivation}, $(k+k')^2\simeq 2k\cdot k'\neq 0$. 
For this reason, $S^{0 \gtrless}(k+k')\simeq 0$, which implies $\VC^{\gtrless}(k,X)\simeq 0$, so  the only nonzero function of the off-diagonal self-energy appearing at Wigner-transformed Eq.~(\ref{eq:app-ana-3}) is $\VC^\R(k,X)\simeq \VC^\A(k, X)$.
Therefore, we drop the second term in Eq.~(\ref{eq:app-ana-3}) because that term becomes negligible after the Wigner transformation, and hence the equation becomes
\begin{align}
\label{eq:app-ana-4}
\begin{split}
&\int_C d^4z  \VC(x,z)D(z,y)
\simeq-i\int^{\infty}_{-\infty} d^4z \VC^\R(x,z) D^<(z,y).
\end{split}
\end{align}

In the same way, we get
\begin{align}
\label{eq:app-ana-5}
\int_C d^4z  S(x,z)\VC(z,y)\simeq i\int d^4z  S^<(x,z)\VC^\R(z,y).
\end{align}
The calculation in QED can be performed in the same way.

Finally we derive Eq.~(\ref{eq:QED-WT2}). 
By multiplying Eq.~(\ref{eq:QED-WT-pre}) by $\int d^4s \exp({ik\cdot s})$, we get
\begin{align} 
\begin{split}
&-\cp\int d^4s e^{ik\cdot s}\Biggl(\frac{\delta{\overline{\eta}_{\text {ind}}}(x)}{\delta\overline{\varPsi}(y)}\varPsi(x)\\
&\qquad-\delta(x^0-y^0)\delta^{(3)}(\vx-\vy)\etaind(x)\Biggr)\\
&\quad=\int d^4s e^{ik\cdot s}\partial^s_\mu \VC^\mu(y,x).
\end{split}
\end{align}
Here we have set $x^0$, $y^0\in C^+$  and neglected the sub-leading terms.
The first term in the left-hand side has the same order of magnitude as the hard fermion self-energy $\varSigma(k)$ times $\varPsi(X)$, so that term is negligible.
Thus the left-hand side becomes $\etaind(X)$.
The right-hand side becomes
\begin{align}
\begin{split}
&\int^0_{-\infty} d^4s e^{ik\cdot s}\partial^s_\mu \VC^{>\mu}(y,x)
+\int^\infty_0 d^4s e^{ik\cdot s}\partial^s_\mu \VC^{<\mu}(y,x)\\
&\quad=\int^\infty_{-\infty} d^4s e^{ik\cdot s}\partial^s_\mu \VC^{>\mu}(y,x)\\
&\qquad+\int^\infty_0 d^4s e^{ik\cdot s}\partial^s_\mu (\VC^{<\mu}(y,x)-\VC^{>\mu}(y,x))\\
&\quad=-ik_\mu \VC^{>\mu}(-k,X)-k_\mu \VC^{\A\mu}(-k,X).
\end{split}
\end{align}
We see that the first term in the last line is negligible because of the on-shell condition.
Thus we obtain Eq.~(\ref{eq:QED-WT2}) if we remember that $\VC^\R(k,X)\simeq \VC^\A(k, X)$.
\section{KINETIC EQUATION IN DIAGONAL CASE}
\label{app:diagonal}
We derive the equation that corresponds to Eq.~(\ref{eq:yukawa-result}) in the diagonal case in this Appendix.
The equation governing the propagator of the fermion is as follows:
\begin{align}
\label{eq:yukawa-diagonal-0}
\begin{split}
& \Slash{D}_x[\varPhi] S(x,y)+i\cp(K(y,x))^\dagger\gamma^0\varPsi(x)\\
& \quad=\delta^{C(4)}(x-y)+i\frac{\delta \etaind(x)}{\delta \eta(y)}.
\end{split}
\end{align}
Since the second term in the left-hand side contains two $\varPsi$, we neglect that term. 
Let us calculate the right-hand side.
We set $x^0\in C^+$ and $y^0\in C^-$.
Since the vertex correction term, which contains more than one $\varPsi$,  is negligible, we obtain
\begin{align}
\begin{split}
\frac{\delta \etaind(x)}{\delta \eta(y)}&= \int _C d^4z \varSigma(x,z) S(z,y)\\
&=-\int^{x^0}_{t^0} d^4z \varSigma^>(x,z) S^<(z,y) \\
&\quad+ \int^{y^0}_{x^0} d^4z \varSigma^<(x,z) S^<(z,y)\\
&\quad-\int^{t^0-i\beta}_{y^0} d^4z \varSigma^<(x,z) S^>(z,y)\\
&=-\int^{x^0}_{t^0} d^4z (\varSigma^>(x,z)+\varSigma^<(x,z)) S^<(z,y)\\
&\quad+ \int^{y^0}_{t^0} d^4z \varSigma^<(x,z) (S^<(z,y)+S^>(z,y)).
\end{split}
\end{align}
By taking the limit $t^0\rightarrow-\infty$, we get
\begin{align}
\begin{split}
&\frac{\delta \etaind(x)}{\delta \eta(y)}\\
&\quad\simeq i\int^{\infty}_{-\infty} d^4z (\varSigma^\R(x,z) S^<(z,y)+ \varSigma^<(x,z) S^\A(z,y)).
\end{split}
\end{align}
Here we have introduced the advanced fermion propagator, $S^\A(x,y)\equiv -i\theta(y^0,x^0)(S^>(x,y)+S^<(x,y))$.

By performing the Wigner transformation, we get 
\begin{align}
\label{eq:yukawa-diagonal-1}
\begin{split}
&\left[-i\Slash{k}+\frac{\Slash{\partial}_X}{2}+i\cp\left(\varPhi(X)-i\frac{\partial_k}{2}\cdot(\partial_X \varPhi(X))\right)\right] S^<(k,X) \\
 &\quad=-i(\varSigma^\R(k,X)S^<(k,X)+\varSigma^<(k,X)S^\A(k,X)).
\end{split}
\end{align}
The following equation is derived from the conjugate of Eq.~(\ref{eq:yukawa-diagonal-1}) by using $\gamma^0S^>(x,y)\gamma^0=S^<(y,x)$:
\begin{align}
\label{eq:yukawa-diagonal-2}
\begin{split}
& S^<(k,X)\left[i\Slash{k}+\frac{\overleftarrow{\Slash{\partial}_X}}{2}-i\cp\left(\varPhi(X)+i\frac{\overleftarrow{\partial_k}}{2}\cdot(\partial_X \varPhi(X))\right)\right]\\
 &\quad=i(S^\R(k,X)\varSigma^<(k,X)+S^<(k,X)\varSigma^\A(k,X)).
\end{split}
\end{align}
Here we have introduced the retarded fermion propagator, $S^\R$, and the advanced fermion self-energy, $\varSigma^\A$, which are defined as follows:
\begin{align}
S^\R(x,y)&\equiv i \theta(x^0, y^0)(S^>(x,y)+S^<(x,y)), \\
\varSigma^\A(x,y)&\equiv -i\theta(y^0,x^0)(\varSigma^>(x,y)+\varSigma^<(x,y)).
\end{align}

By multiplying Eq.~(\ref{eq:yukawa-diagonal-1}) (Eq.~(\ref{eq:yukawa-diagonal-2})) by $-i\Slash{k}+\Slash{\partial}_X/2+i\cp\left(\varPhi(X)-i\partial_k\cdot\partial_X \varPhi(X)/2\right)-i\varSigma^\R(k,X)$ $(i\Slash{k}+\Slash{\partial}_X/2-i\cp\left(\varPhi(X)+i\partial_k\cdot\partial_X \varPhi(X)/2\right)-i\varSigma^\A(k,X))$ from the left (right), we get
\begin{align}
\notag
&(-k^2-ik\cdot\partial_X+\cp(2\Slash{k}\varPhi(X)-i\Slash{k}\partial_k\cdot(\partial_X \varPhi(X))\\
\notag
&\qquad+\varPhi(X)\Slash{\partial}_X)+\{\Slash{k}, \varSigma^\R(k,X)\}) S^<(k,X)\label{eq:yukawa-diagonal-3}\\
 &\quad=-\Slash{k}\varSigma^<(k,X)S^\A(k,X), \\
 \notag
& S^<(k,X)(-k^2+ik\cdot\overleftarrow{\partial_X}+\cp(2\Slash{k}\varPhi(X)+i\overleftarrow{\partial_k}\cdot(\partial_X \varPhi(X))\Slash{k}\\
\notag
&\qquad-i\overleftarrow{\Slash{\partial}_X}\varPhi(X))+\{\Slash{k}, \varSigma^\A(k,X)\})  \label{eq:yukawa-diagonal-4}\\
 &\quad=-S^\R(k,X)\varSigma^<(k,X) \Slash{k} .
\end{align}
By subtracting Eq.~(\ref{eq:yukawa-diagonal-3}) from Eq.~(\ref{eq:yukawa-diagonal-4}), we get
\begin{align}
\label{eq:yukawa-diagonal-5}
\begin{split}
&\left(2ik\cdot\partial_X-\{\Slash{k},\varSigma^R(k,X)\}\right)S^<(k,X)\\
&\qquad+S^<(k,X)\{\Slash{k},\varSigma^\A(k,X)\}\\
&\qquad+2\cp\varPhi(X)[S^<(k,X), \Slash{k}] \\
&\qquad +i\cp(\partial^\nu_X\varPhi(X))\{\Slash{k},\partial_{k\nu}S^<(k,X)\}\\
&\qquad-i\cp \varPhi(X)\{\gamma_\mu, \partial^\mu_X S^<(k,X)\}\\
 &\quad=\Slash{k}\varSigma^<(k,X)S^\A(k,X)-S^\R(k,X)\varSigma^<(k,X)\Slash{k} . 
\end{split}
\end{align} 
Here we linearize this equation.
By introducing $\delta S(k,X)=S(k,X)-S^{0}(k)$ and $\delta\varSigma(k,X)\equiv\varSigma(k,X)-\varSigma^{\text {(eq)}}(k)$, we arrive at the following equation:
\begin{align}
\label{eq:app-diagonal-kinetic}
\begin{split}
&\left(2ik\cdot\partial_X-\{\Slash{k},\varSigma^{{\text {(eq)}}\R}(k)-\varSigma^{{\text {(eq)}}\A}(k)\}\right)\delta S^<(k,X)\\
&\qquad-\{\Slash{k},\delta\varSigma^R(k,X)\} S^{0<}(k)+S^{0<}(k)\{\Slash{k},\delta\varSigma^{\A}(k,X)\}\\
&\qquad+i\cp(\partial^\nu_X\varPhi(X))\{\Slash{k},\partial_{k\nu}S^{0<}(k)\}\\
 &\quad=\Slash{k}\delta \varSigma^<(k,X)S^{0 \A}(k)-S^{0 \R}(k)\delta\varSigma^<(k,X)\Slash{k} \\
 &\qquad+\Slash{k}\varSigma^{{\text {(eq)}} <}(k)\delta S^\A(k,X)-\delta S^\R(k,X)\varSigma^{{\text {(eq)}} <}(k)\Slash{k} . 
\end{split}
\end{align}

Similarly, we  obtain the following equation in QED:
\begin{align}
\label{eq:app-diagonal-kinetic-QED}
\begin{split}
&(2ik\cdot\partial_X-\{\Slash{k},\varSigma^{{\text {(eq)}}\R}(k)-\varSigma^{{\text {(eq)}}\A}(k)\})\delta S^<(k,X)\\
&\qquad-\{\Slash{k},\delta\varSigma^R(k,X)\} S^{0<}(k)+S^{0<}(k)\{\Slash{k},\delta\varSigma^\A(k,X)\}\\
&\qquad+2i\cp k^\mu\partial_k\cdot\partial_X A_\mu(X) S^{0<}(k)\\
&\qquad-\frac{\cp}{2}F^{\mu\nu}(X)[\sigma_{\mu\nu}, S^{0<}(k)]\\
 &\quad=\Slash{k}\delta \varSigma^<(k,X)S^{0 \A}(k)-S^{0 \R}(k)\delta\varSigma^<(k,X)\Slash{k} \\
 &\qquad+\Slash{k}\varSigma^{{\text {(eq)}} <}(k)\delta S^\A(k,X)-\delta S^\R(k,X)\varSigma^{{\text {(eq)}} <}(k)\Slash{k} .
\end{split}
\end{align}

\section{SMALLNESS OF $K^0$}
\label{app:K0}

Here we show that $K^0$ is negligible compared with the spatial components in the Coulomb gauge.
For simplicity, we set $A^\mu=0$.
We get the following equation by multiplying Eq.~(\ref{eq:qed-wigner-fermion}) by $-i\Slash{k}+\Slash{\partial}_X/2+i\varSigma^\R(k,X)$ from the left, subtracting Eq.~(\ref{eq:qed-coulomb-wigner-boson}) from the quantity obtained above, and setting $\mu=0$:
\begin{align}
\begin{split}
(2ik\cdot\partial_X-\{\Slash{k},\varSigma^\R(k,X)\}
&+ (k^0)^2)K_0(k,X)\\
-\varPi^{\A 0\nu}(k,X)K_\nu(k,X)&=\cp S^<(k)\V^0(k,X)\\ 
 (k^0)^2 K_0(k,X)
 -\varPi^{\A 0i}(k,X)&K_i(k,X)\\
 &=\cp S^<(k)\V^0(k,X).
\end{split}
\end{align}
Since $k^0\sim T$, we see that
\begin{align}
K^0\sim \cp^2 K^i.
\end{align}

\section{TEMPORAL GAUGE}
\label{app:temporal}

We show that the equation determining $K^\mu$ in the temporal gauge is the same as that in the Coulomb gauge, Eq.~(\ref{eq:qed-coulomb-result-K}), in this Appendix because the resummed perturbation theory was first proposed in the temporal gauge~\cite{lebedev}.
The gauge-fixing condition in the temporal gauge is $\GF[a]=a^0$ and $\lambda\rightarrow \infty$. 
This condition is equivalent to the constraint $a^0=0$.
Because of this constraint, we have
\begin{align}
K^0(x,y)=D^{0\mu}(x,y)=0.
\end{align}
The equations governing $K^i$ are 
\begin{align}
\notag
&\left(-i\Slash{k}+\frac{\Slash{\partial}_X}{2}+i\cp\Slash{A}(X)\right)K^i(k,X)\\
\label{eq:qed-temporal-wigner-fermion}
&\qquad+i\cp\gamma_j D^{< j i}(k,X)\varPsi(X)\\
\notag
&\quad=-i(\varSigma^\R(k,X)K^i (k,X)+\VC^\R_{j}(k,X) D^{< j i}(k,X)), \\
\notag
&(-k^2+ik\cdot\partial_X)K^{i}(k,X)\\
\label{eq:qed-temporal-wigner-boson}
&\qquad+\left(k-\frac{i\partial_X}{2}\right)^i \left(k-\frac{i\partial_X}{2}\right)^jK_j(k,X)\\
\notag
&\qquad+\cp S^<(k,X)\gamma^i\varPsi(X)\\
\notag
&\quad=\varPi^{\A i j}(k,X)K_j(k,X)-S^<(k,X)\VC^{\R i}(k,X).
\end{align}
From these equations, we obtain
\begin{align}
\label{eq:qed-temporal-K}
\begin{split}
&(-2ik\cdot\partial_X+2\cp k\cdot A(X)+\{\Slash{k},\varSigma^\R(k,X)\})K^i(k,X)\\
&\qquad-\left(k-\frac{i\partial_X}{2}\right)^i \left(k-\frac{i\partial_X}{2}\right)^jK_j(k,X)\\
&\qquad+\varPi^{\A ij}(k,X) K_j(k,X)\\
&\quad=
-(\Slash{k}D^{<  j i} (k,X)+\delta^{ij}S^<(k,X))\V_j(k,X).
\end{split}
\end{align}

Here let us evaluate $k^i K_i(k,X)$, which is the longitudinal component of $K^i(k,X)$.
By multiplying Eq.~(\ref{eq:qed-temporal-K}) by $k_i$, we get
\begin{align}
\begin{split}
&|\vk|^2 \left(k-\frac{i\partial_X}{2}\right)^jK_j(k,X)+k_i\varPi^{\A ij}(k,X) K_j(k,X)\\
&\quad=
- S^<(k,X)k_i\V_i(k,X).
\end{split}
\end{align}
Here we have neglected the terms that are of order $\cp^2 T^2 k^i K_i(k,X)$.
We see that $k^i K_i(k,X)\sim \cp^2T K_i(k,X)$ and thus the longitudinal component of $K_i(k,X)$, $\hat{k}^i K_i(k,X)$, is negligible compared with the transverse component of $K_i(k,X)$.
We note that $K^0=0$, which is the result of the gauge-fixing condition, and $k^i K_i=0$ are valid also in the Coulomb gauge in the leading order.
Furthermore, also the free photon propagator at equilibrium is the same as that in the Coulomb gauge. 
Thus, we can obtain Eq.~(\ref{eq:qed-coulomb-result-K}) in the same way as in the Sec.~\ref{ssc:derivation}.



\begin{thebibliography}{99}
\bibitem{plasmon}
H.~A.~Weldon,
  Phys.\ Rev.\  D {\bf 26}, 1394 (1982).
  
\bibitem{plasmino}
H.~A.~Weldon,
  Phys.\ Rev.\  D {\bf 26}, 2789 (1982);
   {\bf 40}, 2410 (1989).

\bibitem{electron-damping}  
R.~D.~Pisarski,
  Phys.\ Rev.\ Lett.\  {\bf 63}, 1129 (1989);
  J.~P.~Blaizot and E.~Iancu,
  {\it ibid}.
    {\bf 76}, 3080 (1996);
  Phys.\ Rev.\  D {\bf 55}, 973 (1997);
see also 
  V.~V.~Lebedev and A.~V.~Smilga,
  Phys.\ Lett.\  B {\bf 253}, 231 (1991);
  Physica A {\bf 181}, 187 (1992).

\bibitem{HTL}
J.~Frenkel and J.~C.~Taylor,
  Nucl.\ Phys.\  B {\bf 334}, 199 (1990);
    E.~Braaten and R.~D.~Pisarski,
  {\it ibid}. {\bf 339}, 310 (1990).

\bibitem{HTL-resum}
E.~Braaten and R.~D.~Pisarski,
  Nucl.\ Phys.\  B {\bf 337}, 569 (1990).


\bibitem{QED} 
Y.~Hidaka, D.~Satow and T.~Kunihiro,
  Nucl.\ Phys.\  A {\bf 876}, 93 (2012);
   D.~Satow and Y.~Hidaka,
  arXiv:1109.0088 [hep-ph].

\bibitem{blaizot-ultrasoft}
J.~P.~Blaizot and E.~Iancu,
  Nucl.\ Phys.\  B {\bf 557}, 183 (1999);
 {\bf 570}, 326 (2000);
 see also 
    D.~Bodeker,
  Phys.\ Lett.\  B {\bf 426}, 351 (1998);
  Nucl.\ Phys.\  B {\bf 559}, 502 (1999);
   {\bf 566}, 402 (2000);
  Phys.\ Lett.\  B {\bf 516}, 175 (2001);
  Nucl.\ Phys.\  B {\bf 647}, 512 (2002).
  
  \bibitem{lebedev}
 V.~V.~Lebedev and A.~V.~Smilga,
  Annals Phys.\  {\bf 202}, 229 (1990).

\bibitem{susy-kinetic}
 V.~V.~Lebedev and A.~V.~Smilga,
  Nucl.\ Phys.\  B {\bf 318}, 669 (1989).

\bibitem{transport-perturbation}
 S.~Jeon,
  Phys.\ Rev.\  D {\bf 52},  3591 (1995);
   E.~Wang and U.~W.~Heinz,
  Phys.\ Lett.\  B {\bf 471}, 208 (1999);  
    M.~E.~Carrington, D.~f.~Hou and R.~Kobes,
  Phys.\ Rev.\  D {\bf 62}, 025010 (2000);
   M.~A.~Valle Basagoiti,
   {\it ibid}.
  {\bf 66}, 045005 (2002);
   E.~Wang and U.~W.~Heinz,
    {\it ibid}.
    {\bf 67}, 025022 (2003);
   J.~S.~Gagnon and S.~Jeon,
    {\it ibid}.
    {\bf 75}, 025014 (2007) 
  [Erratum-ibid.\  D {\bf 76},  089902 (2007)];
    {\bf 76}, 105019 (2007).
   
\bibitem{transport-kinetic}
  P.~B.~Arnold, G.~D.~Moore and L.~G.~Yaffe,
  JHEP {\bf 0011}, 001 (2000);
   {\bf 0305}, 051 (2003);  
     H.~Defu,
  arXiv:hep-ph/0501284; 

\bibitem{transport-2PI}
     G.~Aarts and J.~M.~Martinez Resco,
 Phys.\ Rev.\  D  {\bf 68}, 085009 (2003);
  JHEP {\bf 0402}, 061 (2004);
   {\it ibid}.
   {\bf 0503}, 074 (2005);  

\bibitem{blaizot-HTL} 
 J.~P.~Blaizot and E.~Iancu,
  Nucl.\ Phys.\  B {\bf 390}, 589 (1993);
  Phys.\ Rev.\ Lett.\  {\bf 70}, 3376 (1993);
  Nucl.\ Phys.\  B {\bf 417}, 608 (1994).

\bibitem{blaizot-review}
 J.~P.~Blaizot and E.~Iancu,
  Phys.\ Rept.\  {\bf 359}, 355 (2002).

\bibitem{lebellac}
M. Le Bellac, ``Thermal Field Theory,'' {\it Cambridge, UK: Univ. Pr. }(1996).

\bibitem{kadanoff-baym}
L.~Kadanoff, G.~Baym, ``Quantum Statistical Mechanics,'' {\it Benjamin, New York} (1962).

\bibitem{scalar}
 M.~H.~Thoma,
  Z.\ Phys.\  C {\bf 66},491 (1995).

\bibitem{background-field-gauge}
B.~S.~De Witt,
 Phys.\ Rev.\ {\bf 162},  1195 (1967), 
 1239 (1967); 
 Phys.\ Rep.\ {\bf 19C}, 295 (1975);
 L.~F.~Abbott,
  Nucl.\ Phys.\  B {\bf 185}, 189 (1981);
   K.~A.~Meissner,
  Acta Phys.\ Polon.\  B {\bf 17} (1986) 409;
 T.~H.~Hansson and I.~Zahed,
  Phys.\ Rev.\ Lett.\  {\bf 58}, 2397 (1987);
  Nucl.\ Phys.\  B {\bf 292}, 725 (1987);
  As a review article, see 
    L.~F.~Abbott,
  Acta Phys.\ Polon.\  B {\bf 13}, 33 (1982).
     
\bibitem{photon-damping} 
 J.~I.~Kapusta, P.~Lichard and D.~Seibert,
  Phys.\ Rev.\  D {\bf 44}, 2774 (1991)
  [Erratum-ibid.\  D {\bf 47}, 4171 (1993)];
   R.~Baier, H.~Nakkagawa, A.~Niegawa and K.~Redlich,
  Z.\ Phys.\  C {\bf 53}, 433 (1992);
   P.~Aurenche, F.~Gelis and H.~Zaraket,
  Phys.\ Rev.\  D {\bf 61}, 116001 (2000);
Phys.\ Rev.\  D {\bf 62}, 096012 (2000);
P.~B.~Arnold, G.~D.~Moore and L.~G.~Yaffe,
  JHEP {\bf 0111}, 057 (2001);
   JHEP  {\bf 0112},  009 (2001); 
 JHEP  {\bf 0206}, 030 (2002).

\bibitem{QGP}
 K.~Yagi, T.~Hatsuda and Y.~Miake,
  ``Quark-gluon plasma: From big bang to little bang,''
  Camb.\ Monogr.\ Part.\ Phys.\ Nucl.\ Phys.\ Cosmol.\  {\bf 23}, 1 (2005).
     
\bibitem{QCD}
 D.~Satow, in preparation.

\end{thebibliography}
\end{document}